\documentclass[a4paper, 10pt, twoside]{article}

\usepackage{geometry}
\usepackage[table]{xcolor}
\usepackage{algorithm}
\usepackage{algpseudocode}

\usepackage{graphicx} 
\graphicspath{{./Figs/}}
\usepackage{subcaption}
\usepackage{multirow}

\usepackage{cellspace}

\usepackage[square,comma,sort&compress]{natbib}
\usepackage{times}
\usepackage{amsmath}
\usepackage{amsthm}
\usepackage{amsfonts}
\usepackage{tikz}
\usepackage[toc,page]{appendix}

\usepackage{hyperref}
\hypersetup{
	colorlinks = false,
	linkcolor=red,          
	citecolor=red, 
}
\usepackage[toc,page]{appendix}
\usepackage{sectsty}

\usepackage[pagewise]{lineno}

\usepackage{cleveref}
\usepackage{appendix}

\usepackage{longtable}

\usepackage{array}

\usepackage{pdflscape}
\usepackage{adjustbox}

\makeatletter
\newcommand\footnoteref[1]{\protected@xdef\@thefnmark{\ref{#1}}\@footnotemark}
\makeatother

\title{Neural networks for option pricing and hedging: \\ a literature review\footnotetext[0]{We thank Agostino Capponi, Marc Chataigner, St\'ephane Cr\'epey, Antoine Jacquier, and Martin Larsson for comments on an early version of this note.}}

\author{Johannes Ruf\thanks{Department of Mathematics, London School of Economics and Political Science. Email: j.ruf@lse.ac.uk} \and Weiguan Wang\thanks{Department of Mathematics, London School of Economics and Political Science. Email: w.wang34@lse.ac.uk}}

\begin{document}

\maketitle

\begin{abstract}
Neural networks have been used as a nonparametric method for option pricing and hedging since the early 1990s. Far over a hundred papers have been published on this topic. This note intends to provide a comprehensive review. Papers are compared in terms of input features, output variables, benchmark models, performance measures, data  partition methods, and underlying assets. Furthermore, related work and regularisation techniques are discussed.
\end{abstract}

\section{Introduction}


Beginning with \cite{malliaris1993neural} and \cite{hutchinson1994nonparametric},  more than one hundred papers in the academic literature concern the use of artificial neural networks (ANNs) for option pricing and hedging. This work provides a  review of this literature. 
The motivation for this summary 
arose from our companion paper \cite{ruf2019neural}. There we continue the discussions of this note; in particular, of  potentially problematic data leakage when training ANNs to historical financial data.

A linear regression model can be thought of as an affine function that maps some input $x$ to an output $y$. Similarly, an ANN can be thought of as a (possibly repeated) composition of linear and nonlinear functions, again mapping some input $x$ to an output $y$. Training an ANN usually corresponds to choosing the linear components so that this mapping is optimal, in some sense, for (a subset of) a given dataset (the {\em training set}) $(x_i, y_i)_i$. Optimality is usually measured by means of a {\em loss function}, which measures the distance between the ANN output and the given data.

The Stone-Weierstrass theorem asserts that any continuous function on a compact set can be approximated by polynomials.  Similarly, the {\em universal approximation}  theorems ensure that ANNs approximate continuous functions in a suitable way. In particular, ANNs are able to capture nonlinear dependencies between input and output.

With this understanding, an ANN can be used for many applications related to  option pricing and hedging. In the most common form, an ANN learns the price of an option as a function of the underlying price, strike price, and possibly other relevant option characteristics. Similarly,  ANNs might also be trained to learn implied volatility surfaces or optimal hedging ratios. In the pricing task, the corresponding loss function is often chosen to be the squared distance of the  observed (simulated) option prices and the ANN predicted prices. In the hedging task, one would compare observed (simulated) option prices and the values of the ANN hedging portfolios.

Let us provide a formal example in the context of the pricing task, namely a two-hidden layer ANN with linear output. Such an architecture maps an input $x$ (usually a vector consisting of several features, such as moneyness, contract-specific implied volatility, etc.) to an output $y$ (the option price) as follows:
\begin{equation*}
y = w_2\cdot \phi(w_1\cdot x).
\end{equation*}
Here $\phi$ is a nonlinear function (the so called \emph{activation function}), 
 $w_1, w_2$ are weight vectors, and the dot denotes the scalar product.
 Training such an ANN corresponds to finding weight vectors $\hat w_1, \hat w_2$ such that the output $\hat y$ of the ANN is close to the option price $y$, for all samples in a subset of the data (the \emph{training set}). As already mentioned, a widely used criterion to measure what 'close' means is the mean squared error.

The papers discussed here mostly study how well such an approximation by an ANN works on either simulated or real datasets. Different performance measures are employed, and often the ANNs are compared to a variety of benchmarks, the simplest one being the Black-Scholes formula. We shall also summarize how the individual papers choose the training data.

The universal approximation theorems allow a `model-based' usage of ANNs. Imagine a data-generating process, along with a computationally involved pricing algorithm, which relies, for example, on solving partial differential equations or Monte-Carlo simulations. When facing such a situation, ANNs can be used to learn directly the pricing formula.  We review this literature in  \Cref{sec:other}.

This paper is organised in the following way.  \Cref{sec:overview} features \Cref{tab:literature-summary},  a summary of the literature that concerns the use of ANNs for nonparametric pricing (and hedging) of options.    \Cref{S:3}   provides a list of recommended papers from  \Cref{tab:literature-summary}.   \Cref{sec:other} provides an overview of related work where ANNs are applied in the context of option pricing and hedging, but not necessarily as nonparametric estimation tools. \Cref{sec:regularization} briefly discusses various regularisation techniques  used in the reviewed literature. 

\section{ANN based option pricing and hedging in the literature} \label{sec:overview}

\cite{bennell2004black}, \cite{chen2012pricing}, and \cite{hahn2013option}\footnote{\cite{hahn2013option} also surveys the use of ANNs to predict realised volatility. Here we do not aim to do so.} provide extensive literature surveys on the application of ANNs to option pricing and hedging problems. Here we complement these surveys with additional and more recent papers.


\Cref{tab:literature-summary} summarises a large part of the literature and compares six relevant characteristics. They are features (or so-called explanatory variables), outputs of the ANN, benchmark models, data partition between training and test sets, and the underlyings along with the time span of the data. In \Cref{tab:literature-summary}, we only list papers that study an ANN's performance for the option pricing and hedging problem with a somehow statistical perspective. Other papers have different approaches, e.g., a computational perspective, and hence do not fit naturally in the table. These papers are  discussed separately in  \Cref{sec:other}.

We have not included a comparison  of methodologies for the parameter estimation or of ANN architectures, such as number of nodes and layers, activation functions, etc. These
specifications  vary strongly between the papers summarized here.  As an overall trend let us only remark that more recent papers use more complex architectures,  in line with improved availability of computational resources.  
We also do not include a paper-by-paper summary of specific conclusions been drawn. However, more than half of the paper abstracts explicitly emphasize the positive performance of ANNs in the option pricing and hedging task. 

Let us explain how to read \Cref{tab:literature-summary}.  It summarises six relevant characteristics that describe how each paper treats the pricing/hedging problem. 
The columns `Features' and `Outputs' show explanatory features given to the ANN as inputs and outputs, respectively.  \Cref{tab:feature-notations} explains notations and abbreviations used for these columns. The `Benchmarks' column lists  non ANN-based techniques with which an ANN is compared. \Cref{tab:benchmarks} explains the corresponding abbreviations. \Cref{tab:performance-measure} presents abbreviations and definitions for the `Performance measures' column, which summarises how  an ANN (and its benchmarks) are evaluated in each paper. The performance measures marked bold  are  related to evaluations along multiple periods. \Cref{tab:index-table} explains abbreviations for the  underlying assets used in each study and listed in the 'Underlyings' column.

Here an `executive summary' of \Cref{tab:literature-summary}:
\begin{itemize}
	\item There exist two  ways of using the stock price and option strike as inputs to an ANN.  Sometimes they are used as two separate features.  Other times, only their ratio (the so-called moneyness) is used as an input.  In the previous ten years, the second approach is used more often.	 See also  Subsection~\ref{sec:features} for a discussion of this point.
	\item There are many different choices of  volatility estimates concerning input features and benchmarks. The conclusions drawn often depend on this choice.    Subsections~\ref{sec:features} and \ref{sec:benchmarks} provide more details on this point. 
	\item Most papers focus on estimating option prices, around fifteen  papers (10\% of all papers listed) on estimating implied volatilities, and very few deal with the hedging problem directly; see also Subsection~\ref{sec:outputs}.
	\item In some studies, data is partitioned into a training and a test set in a way that violates the underlying time series structure. This introduces information leakage and underestimates the generalization error of the ANN. This is further discussed in Subsection~\ref{sec:data-partition}.
\end{itemize}

For the reader interested in a small selection of all these papers, we refer to \Cref{S:3}. 

After reading about 150 papers and creating \Cref{tab:literature-summary}, we would like to offer three pieces of (personal) advice  when implementing ANNs as nonparametric estimation tool of option prices and hedges. First, stationary features should be used as input. Secondly, the ANN performance should be appropriately   benchmarked.  Third, the time series structure should not be violated when partitioning the data set into  training and test sets.

{
	\begin{landscape}
	\newcolumntype{L}[1]{>{\raggedright\let\newline\\\arraybackslash\hspace{0pt}}m{#1}}	
	\rowcolors{2}{gray!25}{white}

	\begin{longtable}[c]{L{12em} L{7em} L{7em} L{7em} L{7em} L{7em} L{12em} }	
		
		\rowcolor{gray!50}
		Authors \& year & Features & Outputs & Benchmarks & Performance measures & Partition method  & Underlyings \\
		\endhead 
		
		\cite{malliaris1993neural,malliaris1993beating} & $S, K$, $\tau$, $\sigma_{\rm IM}$, $r$, lagged $C$ and  $S$ & $C$ & BS-IM & MAE, MAPE, MSE & Chronological & S\&P100. 6M  \\

		\cite{hutchinson1994nonparametric} & $S/K, \tau$ & $C/K$ & BS-H, Linear & \textbf{MATE}, \textbf{PE},  $R^2$  & Chronological & Simulation (BS);\newline S\&P500. 5Y \\ 
			
		\cite{kelly1994valuing} & $S, K, \tau, \sigma_{\rm H}$ & $C$ & CRR &  MAE, \textbf{MTE}, MSE, $R^2$ & ? & Individual stocks. 6M  \\
		
		\cite{boek1995hybrid} & $S/K$, $\tau$,  $\sigma_{\rm H}$, $r$ & $(C - C_{\rm BS-H})/K$ & BS-H & MAPE, $R^2$ & ? & AOSPI. 2Y\\
		
		{\cite{miranda1995intraday}} & ? & $\Delta\sigma_{\rm I}$ & Linear & ? & ? & IBEX35. ? \\
		
		{\cite{krause1996option}} & $C_{\rm BS-H}$, $S$, $K$, $\tau$, $\sigma_{\rm H}$ & $C$ & BS-H & $R^2$ & Chronological & DAX. 3Y\\
		
		{\cite{lachtermacher1996neural}} & $S$, $K$, $\tau$, $\sigma_{\rm H}$, $r$ & $C$ & BS-H & MAE, MAPE, MPE, MSE & Random & Individual stocks. 2M \\

		\cite{lajbcygier1996comparison} & $S/K$, $\tau$, $\sigma_{\rm IH}$ & $(C-C_{\rm BS-IH})/K$   & BS-IH, KR, Linear & MAE, $R^2$ & Chronological & AOSPI. 3Y\\

		\cite{lajbcygier1996comparing}\footnote{\label{ft:missing-paper}We were not able to obtain a copy of this paper.} & ? & ? & BS-?, BW & ? & ? & AOSPI. ? \\
		
		{\cite{lajbcygier1996neural}, \cite{lajbcygier2002comparing}} & $S/K$, $\tau$, $\sigma_{\rm H}$, $r$ & $C/K$ & BS-H, BW, Linear  & MAPE/MAE, MSE, $R^2$ & Random & AOSPI. 2Y \\
			
		{\cite{liu1996option}} & $S$ &  $S$\footnote{The network learns the dynamics of the underlying iteratively and then relies on Monte-Carlo to determine option prices.} & BS-H & MAE, MAX, MSE & Chronological & S\&P500. 5Y\footnote{The network is trained on a five-year long stock price path, but uses only one day's option price data.}\\
		
		\cite{malliaris1996using} & $\tau$, lagged $\sigma_{\rm IM}$, and others & $\sigma_{\rm IM}$ & None & MAE, MSE & Chronological & S\&P100. 1Y\\
		
		\cite{niranjan1996sequential} & $S/K$, $\tau$ & $C/K$ & BS-H & MSE & ? & FTSE100. 11M\\
		
		\cite{qi1996option}\footnote{This paper relies on the PhD thesis \cite{qi1996financial}.} & $S$, $K$, $\tau$, $r$, open interest & $C$ & BS-H & MAE, MSE, $R^2$ & Random & S\&P500. 2M \\
		
		\cite{hanke1997neural} & $S/K$, $\tau$, $\sigma_{\rm G}$,\footnote{Additional GARCH parameters are also added as features.} $r$& $C/K$, $(C-C_{\rm BS-G})/K$ & None & MSE & Chronological & Simulation (SV) \\ 
		
		\cite{herrmann1997neural} & $S, K$, $\tau$, $\sigma_{\rm I}$, $\sigma_{\rm V}$, $r$ & $C$ & BS-V &  MAE, ME, MSE, $R^2$ & ? & Simulation (BS); DAX. 1Y \\
		
		\cite{karaali1997modelling} & $S, K, \sigma_{\rm H}$ & $C$ & None &  None &  Chronological & DEM volatility. 5Y \\
		
		 \cite{lajbcygier1997improved,lajbcygier1997improved2} & $S/K, \tau$, $\sigma_{\rm IH}$ & $(C-C_{\rm BS-IH})/K$ & BS-IH & MAE, SR & Chronological & AOSPI. 1Y \\
		
		\cite{lajbcygier1997pricing}\footnoteref{ft:missing-paper} & $S/K$, ? & $(C-C_{\rm BS-?})/K$ & ? &  ? & ? & AOSPI. ?\\

		\cite{ahmed1998forecasting} & $S/K$, $\tau$, $\sigma_{\rm H}$, volume & $\sigma_{\rm I}$ & None  & MAE, MSE & Random & Individual stocks. 3Y\\
		\cite{anders1998improving} & $S/K$, $S$, $\tau$, $\sigma_{\rm H}$, $\sigma_{\rm V}$, $r$ & $C/K$, $(C-C_{\rm BS-V})/K$ & BS-H, BS-V &   MAE, MAPE,  ME, MSE, $R^2$ & ? & DAX. 3Y \\
		
		\cite{avellaneda1998following} & $S/K, \tau$ & $\sigma_{\rm I}$ & None & \%E & ? & USD-DEM. Several days \\
		
		\cite{garcia1998option,garcia2000pricing} & $S/K$, $\tau$ & $C/K$ & BS-H, Linear & DM, \textbf{MATE}, MSE & Chronological & Simulation (BS);\newline S\&P500. 8Y \\
		
		\cite{white1998genetic} & ?& $C$ & None & MAE, MSE & Random & Simulation (BS)\\

		\cite{chen1999pricing} &  $S, \tau, \sigma_{\rm H}$,  $\Gamma$, $\Delta$, $\rho$, $\mathcal{V}$, volume & $C$ & BS-H, CRR & MAE, MAPE, MSE & Chronological & Individual stocks. 1Y\\

		\cite{geigle1999artificial}\footnote{This paper relies on the PhD thesis \cite{geigle1999thesis}.} & $S/K$, $\tau$, $\sigma_{\rm H}$, $r$ & $C/K$ & BS-H & MAE, MAPE & Chronological & S\&P500. 6Y \\
		
		\cite{hanke1999adaptive} & $S/K$ & $(C-C_{\rm BS-H})/K$ & BS-H & MSE & Chronological & DAX. 1Y\\
		
		\cite{hanke1999neural} & $S/K$, $\tau$, $\sigma_{\rm Cal}$  & $C/K$, $(C-C_{\rm BS-Cal})/K$ & BS-Cal & MSE & Chronological & DAX. 10M \\
		
		\cite{ormoneit1999regularization} & $S/K$ & $C/K$ & BS-H, BS-IH& \textbf{MATE}, MSE, $R^2$ & ? & DAX. 9M\\
		
		\cite{tsaih1999sensitivity} & $S$, $K$, $\tau$, $\sigma_{\rm I}$, $r$ & $C$ & BS-IH & Sensitivity analysis & Chronological & Simulation (BS)\\
		
		\cite{briegel2000dynamic} & $S$, $\tau$ & $C$ & BS-?, lagged $C$ & MSE & ? & FTSE100. 10M \\
		
		\cite{carelli2000profiling} &  $K$, $\tau$ & $\sigma_{\rm I}$ & None & \%E & ? & USD-DEM. Several days\\
		
		\cite{freitas2000sequential,de2000hierarchical} & $S/K$, $\tau$ & $C/K$ & BS-H & $R^2$ & ? & FTSE100. 11M \\
	
		{\cite{galindo2000framework}} & $S$, $K$, $\tau$ & $C$ & Decision tree,  Linear, Nearest neighbour & MSE & ? &Simulation (BS)\\

		\cite{ghaziri2000neural} & $S$, $K$, $\tau$, $\sigma_{\rm H}$, $r$, open interest & $C$ & BS-H & MSE & ? & S\&P500. 2M \\

		\cite{raberto2000learning} & $S/K, \tau$, $|S-K|/\tau$ & $C/K$ & None & None & ? &  BUND. ? \\
		
		{\cite{saito2000neural}\footnoteref{ft:missing-paper}} & ? & ? & BS-? & ? & ? & S\&P500. ? \\

		\cite{white2000pricing} & $S$, $K$, $\tau$, $\sigma_{\rm H}$ & $C$ & BS-H & MAE, MSE & Random & Simulation (BS); \newline Eurodollar. 7M \\
		
		\cite{yao2000option} & $S$, $K$, $\tau$ & $C$ & BS-H &  $R^2$ &  Chronological & NIKKEI225. 1Y \\
		
		 \cite{dugas2001incorporating,dugas2009incorporating} & $S/K, \tau$ & $C/K$ & None &  MSE & Chronological & S\&P500. 5Y \\
		
		\cite{genccay2001pricing} & $S/K, \tau$ & $C/K$ & BS-H & DM, \textbf{MATE}, MSE &  Chronological & S\&P500. 6Y \\
		
		\cite{le2001emulating} & $S$, $K$, $\tau$, $\sigma_{\rm I}$, $r$ & $C$ & None  & MSE & Chronological & Simulation (BS) \\
		
		\cite{meissner2001capturing} & $S/K$, $\tau$, $\sigma_{\rm G}$ & $C/K$ & BS-G & MAE, MAPE, ME, MSE, $R^2$ & ? & Individual stocks. 8M \\
		
		\cite{schittenkopf2001risk} & $\tau$ & Gaussian parameters\footnote{ANNs output parameters for a  Gaussian mixture density as a model for the risk-neutral density.} & BS-H, CS & MAE, \textbf{MATE}, ME, MSE & Chronological& FTSE100. 5Y\\
		
		\cite{andreou2002critical} & $S/K$, $\tau, \sigma_{\rm H}$, $\sigma_{\rm V}$, $r$, and others & $C/K$, $(C-C_{\rm BS-H})/{K}$,  $(C-C_{\rm BS-V})/{K}$& BS-H, BS-V & MdAE & Chronological & S\&P500. 3Y\\
		
		\cite{billio2002option} & $S/K$, $\tau$, $\sigma_{\rm I}$, $r$ & $C/K$ & BS-? & MSE & Chronological & FTSE100. 1Y\\
		
		{\cite{ghosn2002multi}} & $S/K$, $\tau$ & $C/K$ & None & MSE & Chronological & S\&P500. 6Y\\
		
		\cite{healy2002data} & $S$, $K$, $\tau$, $\sigma_{\rm I}$, $r$, spread, open interest, volume & $C$ &  None &   MAE, ME, $R^2$ & Random & FTSE100. 5Y \\

		 \cite{zapart2002stochastic,zapart2003statistical} & Lagged wavelet coefficients & Wavelet coefficients\footnote{An ANN is used to predict the future volatility of the underlying. The volatility is represented in terms of wavelets and the underlying modelled as a binomial tree.} & BS-? & MAE & Chronological & Individual stocks. 6M/1Y\\
		
		\cite{amilon2003neural} & $S/K$, $\tau$, $\sigma_{\rm H}$, $r$, lagged $S$ &  $C_{\rm Ask}/K$, $C_{\rm Bid}/K$ & BS-H, BS-IM & ME, \textbf{MTE}, MSE  & Chronological &  OMX. 2Y\\
		
		\cite{carverhill2003alternative} & $K/S$, $\tau$, $\sigma_{\rm I}$, $r$ & $C/K$, HR & CRR & \textbf{?TE} & Chronological & S\&P500. 11Y \\

		\cite{genccay2003degree} & $S/K, \tau, \sigma_{\rm H}, r $ & $C/K$ & BS-H & DM, MSE &  Chronological  & S\&P500. 6Y\\
		
		 \cite{healy2003confidence,healy2004confidence}\footnote{ These papers also derive prediction intervals  for ANN estimates of option prices. } & $S/K$, $\tau$ & $C/K$ & None & MSE, $R^2$ & Random & FTSE100. 6Y\\
		
		 \cite{lajbcygier2003improving,lajbcygier2004improving} & $S/K, \tau$ & $(C-C_{\rm BS-IH})/{K}$ & None &  MAE, MSE, $R^2$ & Chronological & AOSPI. 3Y \\
		
		{\cite{montagna2003pricing} } & $S$, $\tau$ & $C$ & None & None & ? & Simulation (BS) \\
		
		 \cite{zapart2003beyond}\footnote{This paper also treats the setup of  \cite{zapart2002stochastic}.} & $S/K$, $\tau$, $\sigma_{\rm H}$, $r$ & $C/K$ & BS-? & MAE & Chronological & Individual stocks. ? \\
		
		\cite{bennell2004black} & $S, K, S/K$, $\tau$, $\sigma_{\rm IM}$, open interest, volume  & $C$, $C/K$ & BS-IM & MAE, ME,  MPE, MSE & Chronological & FTSE100. 1Y \\
		
		{\cite{choi2004efficient}} & $S$, $K$, $\tau$, $\sigma_?$ & $C$ & BS-? &? & Random & KOSPI200. 1Y\\
		
		\cite{dindar2004option} & $S, \tau, \sigma_{\rm H}$, $r$ & $K/C$ & None & ? & Random & South Africa Foreign Exchange. 3Y \\
		
		\cite{morelli2004pricing} & $S, K, \tau, \sigma_{\rm I}$, $r$ & $C$ & None & ? & ? & Simulation (BS)\\
		
		\cite{pires2004american,pires2004option} & $K, \tau$, $\sigma_{\rm H}$ & $C$ & SVM & MAX, ME  & ? & Johannesburg Stock Exchange. 3Y \\
	
		 \cite{xu2004barrier} & $S$, $K$, $\tau$, $\sigma_{\rm I}$, $r$ & $C$ & None & $R^2$ & Random & FTSE100. 5Y\\ 
		\cite{charalambous2005hybrid} & $S, K$, $\sigma_{\rm H}$, $r$, correlations\footnote{\label{ft:corr-underlyings}Correlations between underlyings.} & $C - C_{\rm LA}^{10}$ & LA-10 &  MAE, MAX, MSE  & Chronological & Simulation (BS)\\
		
		\cite{hamid2005can} & $S, \tau, \sigma_{\rm H}, r$ & $C$ & None & MAE, MSE & ? & S\&P500. 12Y \\
		
		{\cite{kakati2005pricing}\footnoteref{ft:missing-paper}}  & ? & ? & ? & ? & ? & Individual stocks. ? \\
		
		 \cite{ko2005hedging,ko2009option} & $S$, $K$, $\tau$, $\sigma_{\rm H}$ &  Coefficients\footnote{Coefficients for a linear regression that returns option prices.} & BS-H & \textbf{MATE} & ? & TAIEX. 1Y/2Y\\ 
		
		\cite{lin2005valuation} & $S, K, \tau, \sigma_{\rm H}, r$ & $C$ & BS-H & MAE, MSE & ? &  TAIEX. 2Y \\
		
		\cite{pires2005american} & $K, \tau, \sigma_{\rm H}$ & $C$ & SVM & MAX, ME, MSE & ? & ALSI. 3Y\\
		
		{\cite{tung2005genso}} & $S-K$, $\tau$, $\sigma_{\rm H}$ & $C$ & None & MSE, Correlation\footnote{\label{ft:pearson-correlation}Pearson correlation coefficient, a statistical measure to verify the goodness-of-fit between the predicted and desired function.} & Random & GBP-USD. 1Y \\
		
		\cite{andreou2006robust}\footnote{\label{ft:andreou-thesis}This paper relies on the PhD thesis \cite{andreou2008parametric}.} & $S/K$, $\tau$, $\sigma_{\rm Cal}$, $\sigma_{\rm H}$, $\sigma_{\rm V}$, $r$ &  $C/K$, $(C - C_{\rm BS-Cal})/{K}$, $(C - C_{\rm BS-H})/{K}$, $(C - C_{\rm BS-V})/{K}$ & BS-Cal, BS-H, BS-V & MAE, MSE & Chronological & S\&P500. 3Y\\
		
		\cite{blynski2006comparison} & $S/K, \tau$, $\sigma_{\rm H}$, $\sigma_{\rm IH}$ &  $C/K$, $(C - C_{\rm BS-H})/{K}$, $(C - C_{\rm BS-N})/{K}$ & BS-H, BS-IH
		& MAE, MAPE, ME, MSE, $R^2$ & ? & S\&P100. 7Y\\
		
		 \cite{huang2006hybrid}, \cite{huang2008online} & $S/K$, $\tau$, $\sigma_{\rm K}$ &  $(C - C_{\rm BS-K})/K$ & SVM & MAE, MAPE, MSE & Chronological & TAIEX. 9M \\
		
		 \cite{jung2006novel} & $S$, $K$, $\tau$,  $\sigma_{\rm IH}$ & C &  BS-IH & MSE & ? & KOSPI200. 1Y\\
			
		 \cite{kim2006local} & $K$, $\tau$ & $\sigma_{\rm Cal}$ & SI & MSE & ? & S\&P500. 1M\\
		
		\cite{liang2006pricing} & $\hat{C}$\footnote{\label{ft:liang}Various price estimations from parametric option pricing models.} & $C$ &  BS-?, CRR & MAE & Chronological & Individual stocks. 5M\\
		
		\cite{mitra2006improving} & $S, K, \tau, \sigma_{\rm H}$, $r$ & $C$ & None &  MAE, MSE & Chronological & NIFTY50. 1Y \\
		
		\cite{pande2006new} & $S/K$, $\tau$, $\sigma_{\rm PCA}$, $r$ & $C$ or(?) $C/K$  & None &  ME, MSE, Correlation\footnote{Correlation between the actual and computed prices.} & ? & Individual stocks. 1Y\\
		
		{\cite{teddy2006brain}} & $S-K$, $\tau$, $\sigma_{\rm H}$ & $C$ & None & MSE, Correlation\footnoteref{ft:pearson-correlation} & Random & GBP-USD. 1Y \\
		
		\cite{tzastoudis2006improving} & $S, K, \sigma_{\rm H}$ & $C$ & BS-H & MAE, $R^2$ & Chronological & S\&P500. Several days \\
		
		{\cite{wang2006dual}} & $S/K$, $\sigma_{\rm IH}$, $(S-K)^+$, $C-(S-K)^+$, $C S/\sqrt{K}$ & $\sigma_{\rm I}$ & BS-IH & MAE, MSE, $R^2$ & ? & Individual stocks. 2M\\

		\cite{amornwattana2007hybrid} & $S$, $K$, $\tau$, $r$ &  $C - C_{\rm BS-N}$, $\sigma_{\rm I}$ & BS-H, BS-N & MAE, MSE& Chronological & Individual stocks. 3M \\
		
		\cite{genccay2007model} & $S$, $K$, $\tau$, $\sigma_{\rm G}$, $r$ & $C$ &  BS-G, BS-H, SV, SVJ & MAE, MSE & ? & S\&P500. 3Y \\
		
		\cite{gregoriou2007hedging} &  $S_{\rm Ask}$, $S_{\rm Bid}$, $S_{\rm Mid}$, $K$, $\tau$,  $\sigma_{\rm I}$, $r$ & $C$ & None & None &  Random & FTSE100. 5Y\\
		
		{\cite{healy2007non}} & $S$, $K$, $\tau$, $\sigma_{\rm I}$, $r$ & $C$ & None & $R^2$ & Chronological & FTSE100. ?\\

		\cite{thomaidis2007comparison} & $S, K, \tau$ & $C$ & BS-G, BS-H &  MAE, MSE &  Chronological & S\&P500. Several days \\
		
		\cite{zhou2007nonparametric} & $S/K$, $S$, $K$, $\tau$, $r$ & $C/K$ & BS-?, CRR & MAE, MAPE, ME, MSE, $R^2$     &  Chronological &Convertible bonds. 2Y \\

		\cite{andreou2008pricing}\footnoteref{ft:andreou-thesis} & $S/K$, $\sigma_{\rm Cal}$, $\sigma_{\rm H}$, $\sigma_{\rm V}$, $r$, kurtosis, skewness &   $C/K$, $(C-C_{\rm BS-Cal})/{K}$, $(C-C_{\rm BS-H})/{K}$, $(C-C_{\rm BS-V})/{K}$ &   BS-Cal, BS-H, BS-V, CS &   MAE, {\textbf{MATE}}, MdAE, MSE, \textbf{MTE} & Chronological & S\&P500. 4Y\\ 
		
		\cite{chiu2008exploring} & $S$, $C_{\rm BS}$, volume, and others & $C$ & None & MSE &  Chronological & Individual stocks. 1Y\\

		{\cite{kakati2008option}} & $S/K$, $\tau$, $\sigma_{\rm G}$, $\sigma_{\rm H}$, $\sigma_{\rm IH}$, $r$ & $C/K$ & BS-G, BS-H, BS-IH & MSE & ? & Individual stocks. Several days\\
		
		\cite{mostafa2008neural}\footnote{This paper relies on the PhD thesis \cite{mostafa2011application}.} & $S/K, \tau, \sigma_{\rm H}$ & $C/K$, $\sigma_{\rm I}$ &  BS-H, SV &  MAPE, \textbf{MATE}, MPE & ? & FTSE100. 2Y \\
		
		 \cite{quek2008novel} & lagged $C$ & $C$ & None & None & ? &  GBP-USD, Gold, Oil. 2Y \\
		
		\cite{saxena2008valuation} & $S/K$, $\tau$, $\sigma_{\rm H}$, $r$ & $(C-C_{\rm BS-H})/{K}$ & BS-H  &  MAE, ME, MPE, MSE, $R^2$   & ? & NIFTY50. 1Y\\
		
		 \cite{teddy2008cerebellar} & $S-K$, $\tau$, $\sigma_{\rm H}$ & $C$ & BS-H & MSE, $R^2$ & Random & GBP-USD. 9M \\ 
		
		\cite{tseng2008artificial} & $S, K, \tau$, $\sigma_{\rm G}$, $r$ & $C$ & None &  MAE, MAPE, MSE  & ? & TAIEX. 2Y \\
		
		{\cite{chen2009learning}}  & $S$, $K$, $\tau$, $\sigma_{\rm H}$, $r$ & $C$ & BS-H, SVM & MAE, MSE & Chronological & S\&P500. Several days \\

		\cite{gradojevic2009option} & $S/K, \tau$ & $C/K$ & BS-H & DM, MSE, MSPE  & Chronological &S\&P500. 8Y \\
		
		 \cite{leung2009making} & $\sigma_{\rm H}$, $\sigma_{\rm IH}$, volume, open interest & $\sigma_{\rm I}$ & BS-IH, Linear, Polynomial & ME & Chronological & Several currencies. 17Y\\
		
		\cite{liang2009improving} & $\hat{C}$\footnoteref{ft:liang} & $C$ & CRR, SVM & MAE, MAPE & Chronological & Individual stocks. 2Y\\
		
		\cite{martel2009financial} & $S/K, \tau, \sigma_{\rm H}$, $r$ & $C_{\rm Bid}/K$, $C_{\rm Ask}/K$ & BS-H & ME, MSE, \textbf{MTE} & Chronological &  IBEX35. 2Y \\

		\cite{samur2009use} & $S, K, \tau, \sigma_{\rm H}$, $r$ & $C$ & None & MAE, MSE, $R^2$ & ? & S\&P100. Several days \\
		
		\cite{wang2009nonlinear} & $S/K$, $\tau$, $\sigma_{\rm G}$, $r$ & $C/K$ & None & MAE, MAPE, MSE & ? & TAIEX. 2Y \\ 
	
		{\cite{wang2009using}} & $S/K$, $\tau$, $\sigma_{\rm G}$, $\sigma_{\rm H}$, $\sigma_{\rm IH}$, $r$ & $C/K$ & None & MAE, MAPE, MSE & ? & TAIEX. 2Y\\
		
		\cite{andreou2010generalized}\footnoteref{ft:andreou-thesis} & $S/K, \tau$ & $\sigma_{\rm I}$ & BS-Cal, CS, {SV, SVJ} &  MAE, \textbf{MATE}, MdAE, MSE  & Chronological & S\&P500. 3Y\\ 
		
		\cite{barunikova2011neural} & $S, K, \tau$ & $C$ & BS-H & MAE, MAPE, MSE   & Random & S\&P500. 3Y \\
		
		\cite{gradojevic2011parametric} & $S/K$, $\tau$, $\sigma_{\rm IH}$, $r$ & $C/K$ & BS-H & DM, MAPE, MSE & Chronological & S\&P500. 7Y\\
		
		\cite{liu2011pricing} & $S/K, \tau$, $\sigma_{\rm H}$,\footnote{More precisely, a Markov regime switching model is used to estimate the volatility.} $r$ & $C/K$ & BS-H &  MAE, MSE & Chronological & Individual stocks. 2Y\\
		
		\cite{phani2011quest} & $S, K, \tau$ & $C$ & BS-?, SVM &MAE & ? & NIFTY50. 2Y \\
		
		{\cite{tung2011financial}} & $\sigma_{\rm IH}$ & $\sigma_{\rm I}$ & None & MAPE, MSE, $R^2$ & Chronological & HSI. 5Y \\
		
	 	\cite{wang2011pricing} & $S/K$, $S$, $\tau$,  $\sigma_{\rm Cal}$, $r$ & $C$ & SV, SVJ, SVM &  MAE, MAPE & Chronological & Several currencies.  7M\\
		
		 \cite{ahn2012applying} & Lagged $\sigma_{\rm I}$, Greeks &  $\text{Sign}(\Delta \sigma_{\rm I})$ & None & Accuracy & 
		Chronological & KOSPI200. 2Y\\

		\cite{chen2012pricing} & $S/K, \tau$ & $C/K$, $(C-C_{\rm BS-H})/K$, HR & BS-H &  MAE, ME, MSE  & Random & Sterling futures. 2Y \\
		
		\cite{mitra2012option} & $S$, $K$, $\tau$, $\sigma_{\rm H}$, $r$ & $C$ & BS-H & ME, MSE & Chronological & NIFTY50. 3Y \\
		
		\cite{shin2012dynamic} & $S, K, \tau$, $r$ & HR & None & MPE & Chronological & KOSPI200. 10Y\\	
		
		 \cite{wang2012using} & $S$, $K$, $\tau$, $\sigma_{\rm Cal}$,  $\sigma_{\rm G}$, $\sigma_{\rm H}$, $\sigma_{\rm IH}$& $C$ & None & MAE, MAPE, MSE & Chronological & TAIEX. 2Y \\
		
		\cite{chang2013forecasting} & $S/K, \tau, \sigma_{\rm G}$, $r$ & $C$ or(?) $C/K$ & None & MAE, MAPE & ? & TAIEX. 2Y\\
		
		 \cite{hahn2013option} & $S/K$, $\tau$, $\sigma_{\rm H}$, $r$ & $C/K$ & SV &  MAE, MAPE, MSE & Chronological & Individual stocks. 10Y\\
		
		\cite{can2014nonparametric} & $S/K, S, \tau$, $r$ & $C/K$ & BS-H & MAE & Chronological &S\&P100. Several days \\

		\cite{lai2014comparison} & $S/K$, $\tau$, $r$ & $\sigma_{\rm I}$ & KR, SI & KS & ? & Simulation (BS, SV, SVJ)\\
		
		\cite{park2014parametric} & $S/K, \tau$ & $C/K$ & BS-H, SV & MSE & Chronological & KOSPI200. 10Y\\
		
		\cite{von2014steps} & $S/K$, $K$, $\tau$ & $C/K$ & None & MSE, $R^2$ & Chronological & EUR-USD. 1M \\
		
		\cite{ludwig2015robust} & $S/K, \tau$ & $\sigma_{\rm I}$ & Quadratic & MSE, $R^2$ & ? & S\&P500. 12Y\\

		\cite{liu2016performance} & $S/K, \tau$ & $(C-C_{\rm BS-H})/{K}$ & BS-H &  MAE, MAPE, ME, MSE   & ? & HSI. 6Y \\
		
		\cite{montesdeoca2016extending} & $S/K, \tau$, $\sigma_{\rm H}$, volume & $C/K$ & None & MSE & Chronological & FTSE100. ?; \newline Individual stocks. ? \\ 
		
		\cite{culkin2017machine} & $S/K, \tau, \sigma_{\rm I}$, $r$ & $C/K$ & None &  MSE, $R^2$&  Chronological & Simulation (BS)\\
		
		 \cite{das2017new} & $S/K$, $\tau$,  $\widehat C$\footnoteref{ft:liang} & $C$ & BS-H, SVM & MAE, MSE & Chronological & NIFTY50. 2Y\\
		
		\cite{fang2017application} & $\sigma_{\rm H}$ & $\sigma_{\rm I}$ & None &  MSE, $R^2$ & Chronological & Simulation (BS); WTI. 1M\\
		
		 \cite{palmer2017pseudo} & $S$, $K$, $\sigma_{\rm I}$, $r$ & $C$ & None & MAE, MdAE, MAPE & Chronological & Simulation (BS) \\

		\cite{yang2017gated}\footnote{This paper relies on the PhD thesis \cite{zheng2017machine}.} & $K/S, \tau$ & $C/S$ & BS-?, Kou, VG  & MAPE, MSE & ? & S\&P500. 10Y \\

		\cite{ferguson2018deeply} & $S$, $\tau$, $\sigma_{\rm I}$, correlations\footnoteref{ft:corr-underlyings} & $C$ & None & MSE  & Chronological & Simulation (BS) \\

		\cite{ackerer2019deep} & $\log(K/S), \tau$, $\log(K/S)\tau^{-0.5}$, $\log(K/S)\tau^{-0.95}$ & $\sigma_{\rm I}$ &  None &  MAPE, MSE & Random &  S\&P500. 1M\\
		
		\cite{buehler2019deep,buehler2019deep-ri} & $\log(S)$ & HR & BS-I & \textbf{CVaR} & Chronological & Simulation (BS, SV); S\&P500. 5Y\\
		
		\cite{cao2019neutral} & $S/K, \tau$,  $\sigma_{\rm V}$, underlying return & $\sigma_{\rm I}$ & HW & MSE &  Random & S\&P500. 8Y\\


		\cite{jang2019generative} & ? & $C$  &  BS-Cal, BW, KR, LSM, LV,  SVJ, SVM   & MAE, MAPE, MPE, MSE & ? & S\&P100. 9Y\\
		
		\cite{liu2019pricing} & $S/K, \tau$ & $\sigma_{\rm I}$ &  None & MAE,  MAPE, MSE & Chronological & Simulation (BS)\\

		\cite{liu2019wavelet} & $S/K, \tau$, $\sigma_{\rm Cal}$, $r$ & $(C-C_{\rm BS-H})/{K}$ &  BS-Cal, SVJ &  MAE, \textbf{MATE}, MPE,  MSE& Chronological & DAX. 4Y \\
		
		\cite{karatas2019supervised} & $S/K$, $\tau$, $r$, ? & $C/K$ & None & MSE, $R^2$ & Chronological & Simulation (BS, SV, VG)\\
		
		\cite{palmer2019evolutionary} & $S/K$, $\sigma_{\rm I}\sqrt{\tau}$, $r$ & $C/K$ & BS-I, LSM & MAE, MAPE & Chronological & Simulation (BS) \\

		\cite{zheng2019gated} & $S/K, \tau$ & $\sigma_{\rm I}$ & SSVI & MAPE & ? & S\&P500. 10Y \\	
		
				\cite{ruf2019neural} & $S/K$, $\sigma_{\rm I}\sqrt{\tau}$, $\Delta$, $\mathcal{V}$, Vanna
		& HR & BS-I, HW, Linear & MSE & Chronological & Simulation (BS, SV); S\&P500. 8Y; \newline STOXX50. 3Y  \\
	
		\hline
		
		\rowcolor{white}
		
		\caption{This table summarises more than 150 papers that use ANNs as a nonparametric option pricing or hedging tool.  These papers are compared in terms of features (or so-called explanatory variables), outputs of the ANN, benchmark models, data partition between training and test sets, and the underlyings along with the time span of the data. The performance measures marked bold are  related to evaluations along multiple periods. We refer to Tables~\ref{tab:feature-notations}--\ref{tab:index-table} for a dictionary of all abbreviations used here.}	
		\label{tab:literature-summary}
		
	\end{longtable}
\end{landscape}
}

\begin{table}
	\centering
	\begin{tabular}{m{5em} m{30em}}
		\hline
		$C$ & Option price \\
		$C_{\rm BS-X}$ & Option price given by the Black-Scholes formula; see \Cref{tab:benchmarks} for the different meanings of X\\
		$C_{\rm LA}^{n}$ & Option price given by $n$-step multi-dimensional lattice scheme \\
		HR & Hedging ratio \\
		$K$ & Strike price \\
		$S$ & Stock price \\
		$r$ & Interest rate \\

		$\Gamma$ & Gamma: second-order sensitivity of option price with respect to underlying price \\
		$\Delta$ & Delta: sensitivity of option price with respect to underlying price \\
		$\mathcal{V}$ & Vega: sensitivity of option price with respect to volatility \\
		
		$\rho$ & Rho: sensitivity of option price with respect to interest rate \\
		$\sigma_{\rm Cal}$ & Volatility from calibration (e.g., constant across strikes and maturities) \\
		$\sigma_{\rm G}$ & GARCH--generated volatility \\
		$\sigma_{\rm H}$ & Historical volatility \\
		$\sigma_{\rm I}$ & Implied volatility \\
		$\sigma_{\rm IH}$ & Implied historical volatility \\
		$\sigma_{\rm IM}$ & At-the-money implied  volatility \\
		$\sigma_{\rm K}$ & Volatility obtained from Kalman filter \\
		$\sigma_{\rm PCA}$ & Macroeconomic variables that contribute the most to volatility, determined by principle component analysis\\
		$\sigma_{\rm V}$ &  Volatility index such as VIX and DVAX \\
		$\tau$ & Time to maturity \\
		\hline
		
	\end{tabular}
	\caption{This table presents notations and abbreviations for features and outputs, used in \Cref{tab:literature-summary}.}
	\label{tab:feature-notations}
\end{table}

\begin{table}
	\centering
	\begin{tabular}{m{5em} m{30em}}
		\hline
		
		BS-Cal & Black-Scholes formula with calibrated volatility \\ 

		BS-G & Black-Scholes formula with GARCH-generated volatility \\
		BS-H & Black-Scholes formula with historical volatility \\
		
		BS-I & Black-Scholes formula with contract-specific implied volatility\\
		
		BS-IH & Black-Scholes formula with historical implied volatility \\
		BS-IM & Black-Scholes formula with at-the-money implied volatility \\
		
		BS-K & Black-Scholes formula with  volatility obtained from Kalman filter \\ 
		
		BS-N & Black-Scholes formula with ANN-generated volatility \\
		BS-V & Black-Scholes formula with volatility index, such as VIX or VDAX\\
		BW & \cite{barone1987efficient} pricing method \\
		CRR & \cite{cox1979option} model \\
		CS & \cite{corrado1996skewness} model \\
		
		HW & \cite{hull2017optimal} model \\
		Kou  & \cite{kou2002jump}'s jump diffusion model\\
		KR & Kernel regression \\
		
		LA-n & n-step multi-dimensional lattice scheme \\
		Linear & Linear regression on features\\
		LSM & \cite{longstaff2001valuing} method \\
		LV & Local volatility model \\
		Quadratic & Quadratic regression on features\\
		SI & Spline interpolation \\
		SSVI & Surface stochastic volatility inspired model, see \cite{gatheral2014arbitrage} \\
		
		SV &  Stochastic volatility models, such as \cite{heston1993closed} or GARCH  \\
		SVJ & Stochastic volatility with jumps model, see \cite{bates1996jumps} or \cite{carr2003stochastic} \\
		SVM & Support vector machine \\
		
		VG & Variance Gamma model, see \cite{madan1998variance} \\
		
		\hline	
	\end{tabular}
\caption{This table presents abbreviations for various benchmarks, used in \Cref{tab:literature-summary}.}
\label{tab:benchmarks}
\end{table}

\clearpage
	\begin{longtable}[c]{m{5em} m{15em} m{15em} }
		
		\hline
		
		DM & Diebold and Mariano test & \\
		
		KS & Kolmogorov and Smirnov two-sample test & \\
		
		MAE & Mean absolute error & \begin{equation*}
		\frac{1}{N}\sum|\hat{y}_i - y_i|
		\end{equation*} \\
		MAPE & Mean absolute percentage error & \begin{equation*}
		\frac{1}{N}\sum\frac{|\hat{y}_i - y_i|}{y_i}
		\end{equation*} \\
		MAX & Maximum error & \begin{equation*}
		\max_i|\hat{y}_i - y_i|
		\end{equation*}\\
		MdAE & Median absolute error  & \begin{equation*}
		\sup_z  \left\lbrace \frac{1}{N}\sum\mathbf{1}_{\left|\hat{y}_i - y_i\right| < z} \leq 0.5 \right\rbrace 
		\end{equation*}\\
		ME & Mean error &  \begin{equation*}
		\frac{1}{N}\sum(\hat{y}_i - y_i)
		\end{equation*}\\
		MPE & Mean percentage error & \begin{equation*}
		\frac{1}{N}\sum\frac{\hat{y}_i - y_i}{y_i}
		\end{equation*}\\
		MSE & Mean squared error & \begin{equation*}
		\frac{1}{N}\sum(\hat{y}_i - y_i)^2
		\end{equation*}\\
		
		
		$R^2$ & Coefficient of determination & \begin{equation*}
		1-\frac{\sum(\hat{y}_i - y_i)^2}{\sum(\bar{y} - y_i)^2}
		\end{equation*}\\

		SR & Sharpe ratio of a trading ratio & \\
		\%E & Sample-wise percentage error & \begin{equation*}
		\frac{\hat{y}_i - y_i}{y_i}
		\end{equation*}\\

		\hline
		
		\textbf{CVaR} & Conditional value-at-risk & \\

		\textbf{MATE} & Mean absolute tracking error & \begin{equation*}
		\frac{1}{N}\sum e^{-rT_i}|V(T_i)|
		\end{equation*}\\
		
		\textbf{MTE} & Mean tracking error & \begin{equation*}
		\frac{1}{N}\sum e^{-rT_i}V(T_i)
		\end{equation*}\\

		\textbf{PE} & Prediction error & \begin{equation*}
		\sqrt{{\rm MTE}^2 + \frac{1}{N}\sum\left(e^{-rT_i}V\left(T_i\right) - {\rm MTE}\right)^2}
		\end{equation*}\\
		\hline
		~\\
\caption{This table presents abbreviations and definitions for performance measures, used in \Cref{tab:literature-summary}. Here, $\hat{y}_i$ is the estimated option price / implied volatility / portfolio value, $y_i$ is the target value, $\bar{y}$ is the average of target values, and $N$ denotes the number of samples. Moreover, $V(T)$, also called tracking error, denotes the terminal value at $T$ of a hedged option portfolio starting with zero wealth. All performance measures marked   bold  are  related to evaluations along multiple periods.}		
		\label{tab:performance-measure}
	\end{longtable}
\clearpage

\begin{table}
	\centering
	\begin{tabular}{m{5em} m{30em}}
		\hline
		
		ALSI & South African All Share Index \\
		AOSPI & Australian  All Ordinaries Share Price Index \\
		
		BUND & German treasury bond \\
		DAX & German stock index\\
		DEM & Deutsche Mark \\
		FTSE100 & UK Financial Times Stock Exchange 100 index \\
		HSI & Hong Kong Heng Seng Index\\
		IBEX35 & Spanish stock index\\
		
		KOSPI200 & Korea Composite Stock Price Index \\
		
		NIFTY50 & Indian National Stock Exchange Fifty \\
		NIKKEI225 & Japanese stock index\\
		
		OMX & Swedish stock index \\

		S\&P100 & US Standard \& Poor's 100\\
		S\&P500 & US Standard \& Poor's 500\\
		STOXX50 &  Eurozone stock index \\
		TAIEX & Taiwanese stock index \\
		WTI & US Light Sweet Crude Oil Futures \\
		\hline
	\end{tabular}
	\caption{This table presents abbreviations for various stock market indices and other underlyings, used in \Cref{tab:literature-summary}. For the shortcuts used to describe simulation data, we refer to \Cref{tab:benchmarks}.}
	\label{tab:index-table}
\end{table}

 \label{sec:category}
In the following, we compare and classify papers listed in \Cref{tab:literature-summary} in terms of features, outputs, performance measures and benchmarks, data partition methods, underlying assets and time span.

\subsection{Features} \label{sec:features}
	To estimate the option price, the underlying price and the strike price are two indispensable variables. Two ways of feeding these two variables into an ANN as input have been suggested.    One way is to use the  underlying price and strike price separately. An alternative is to use a ratio (i.e., moneyness) instead. Several arguments are formulated in the literature in favor of using moneyness:
	\begin{itemize}
		\item Using moneyness instead of the stock price and the strike price separately reduces the number of inputs and thus makes the training of the ANN easier; see \cite{hutchinson1994nonparametric}.
		\item Many parametric models assume that the statistical distribution of the underlying asset's return is independent of the level of the underlying. Hence, the option pricing function is homogeneous of degree one with respect to the underlying stock price and the strike price, so that only moneyness is needed to learn the function. Incorporating this assumption into the ANN can potentially reduce overfitting; see \cite{hutchinson1994nonparametric}, \cite{lajbcygier1997improved, lajbcygier1997improved2}, \cite{anders1998improving}, and \cite{garcia1998option,garcia2000pricing}. 
		\item  Moneyness is a stationary input feature in contrast to the stock price and the strike price. Using it helps generalisation and reduces overfitting; see \cite{ghysels1997nonparametric} and \cite{garcia1998option,garcia2000pricing}.  Our own experiments also confirm that the use of moneyness can significantly improve the generalisation.
	\end{itemize}
\cite{bennell2004black} undertake a systematic experiment on various choices of input features, including underlying price, strike price, moneyness, and on choices of outputs, including option price and option price divided by strike.

Apart from the underlying price and the strike price, volatilities are also widely used as input features. This can be done in several different ways. The most relevant ones are the following:
\begin{itemize}
	\item Using historical volatility estimates as features.
	\item Using volatility indices such as VIX as features.
	\item Using implied volatilities as features.
	\item Using  GARCH forecasts of  (realised or implied) volatility as features.
\end{itemize}
 \Cref{tab:feature-notations} lists further  volatility features.  
The choices of features by the different papers are worked out  in the `Features' column of \Cref{tab:literature-summary}. There exist also several papers that do not use any volatility-type feature as input for their ANNs. 

A few papers, e.g., \cite{blynski2006comparison}, \cite{andreou2008pricing}, or \cite{wang2009using},  compare different volatility features.  Here we summarize their results. \cite{blynski2006comparison} show an ANN outperforms the conventional Black-Scholes when using historical volatility as input, but underperforms when using implied volatility. \cite{andreou2008pricing} show that replacing historical by implied volatility improves the performance of ANNs. \cite{wang2009nonlinear} argue that an ANN with a GARCH volatility forecast    outperforms that with historical and implied volatility as features.

Some papers investigate whether additional features can help the ANN with prediction. To name a few,  \cite{ghaziri2000neural} and \cite{healy2002data}  incorporate option open interests. \cite{samur2009use}  study whether the inclusion of variance improves the performance of the ANN.  \cite{montesdeoca2016extending} explore the potential prediction power of trading volume, option interest, and other variables. \cite{cao2019neutral} investigate the benefit from using the underlying return.

\subsection{Outputs}  \label{sec:outputs}
	
The papers of \Cref{tab:literature-summary} can also be categorised in terms of their outputs:
\begin{itemize}
	\item The most common output is the option price. Depending on whether moneyness is used,  or  underlying price and  strike price are used separately, the output can be the option price or the option price divided by the strike price. 
	Some papers also investigate   the ANN's ability when it is trained to learn the so-called bias; i.e., the difference between market price and a price estimated by a parametric model. Such an ANN is called  hybrid ANN; see, for example, \cite{boek1995hybrid} or \cite{lajbcygier1997improved, lajbcygier1997improved}.
While most of the early papers train their ANNs to fit prices, \cite{garcia2000pricing} train  to  prices, but validate to hedging errors in order to determine the network size that gives the lowest hedging error.  \cite{andreou2010generalized} emphasize the relevance of choosing the right loss function when interested in the hedging  task.
		
	\item Another type of output is the implied volatility. The obtained implied volatilities can be converted to option prices by the Black-Scholes formula. \cite{mostafa2008neural} compare ANNs that output option prices to ANNs that output implied volatilities.  More recently, \cite{liu2019pricing} evaluate an ANN's ability to approximate the  inverse  of the Black-Scholes formula. 
	
	\item The third kind of output (always denoted by HR in \Cref{tab:literature-summary}) is a sensitivity or a hedging ratio.  Only a few papers discuss such an architecture for an ANN.  The first papers are \cite{carverhill2003alternative}, \cite{chen2012pricing}, and \cite{shin2012dynamic}. 
	More recently, \cite{buehler2019deep,buehler2019deep-ri}  and \cite{ruf2019neural} follow up on this line of research.  \cite{buehler2019deep-ri} consider also the hedging of exotic options such as barrier options.
\end{itemize} 

We could have also added the so-called calibration papers to \Cref{tab:literature-summary}, which construct ANNs to map prices to specific model parameters or vice versa.  Instead we decided to dedicate \Cref{SS:calibration} below to these papers.

\subsection{Performance measures and benchmarks} \label{sec:benchmarks}

When evaluating the performance of ANNs, common statistical measures are mean absolute error (MAE), mean absolute percentage error (MAPE), and mean squared error (MSE).\footnote{Several papers use equivalent versions of the measures in Table 4. For example, sometimes root mean squared error is used instead of mean squared error. For consistency, in Table 1, we have made the corresponding adjustments.} These are related to evaluations over a single period, in terms of pricing or hedging. Some papers also propose to evaluate the ANN's performance over multiple periods. For instance, \cite{hutchinson1994nonparametric} introduce the mean absolute tracking error  (MATE) and prediction error (PE), which appear also in many later papers. \cite{buehler2019deep} introduce the conditional value-at-risk (CVaR) for evaluating hedging strategies. 

An ANN's performance should also be compared to a benchmark, for example, a parametric pricing model. The most widely used benchmark is the Black-Scholes formula, which requires a volatility as input. As \Cref{tab:literature-summary} summarises a historic volatility estimate is  used the most often. Also certain implied volatilities (e.g., historical or at-the-money)  appear in the literature.
\cite{blynski2006comparison} compare historical realised and historical implied volatility for the Black-Scholes benchmark.

 The Black-Scholes formula with contract-specific implied volatility  is a valid benchmark for the hedging task.  For the  pricing task, however, such a benchmark would lead to zero error as by definition of implied volatility it prices options without errors. Thus, for the pricing task, the Black-Scholes formula with contract-specific implied volatility
 is not a suitable benchmark .

In addition to the Black-Scholes formula, other widely used parametric benchmarks are   stochastic volatility pricing models; e.g., used in \cite{genccay2007model}, \cite{jang2019generative}, or \cite{liu2019pricing}. 
 \cite{ruf2019neural} observe that if a benchmark is chosen that incorporates both delta and vega hedging then an ANN does not outperform even a simple two-factor regression model.

For American type options,  benchmarks used are the \cite{barone1987efficient}  pricing method (e.g., \cite{lajbcygier2002comparing}), and the Cox-Ross-Rubinstein model (e.g.,  \cite{chen1999pricing}).

\subsection{Data partition methods} \label{sec:data-partition}

An ANN needs to be trained on a training set (in-sample) and then tested on a test set (out-of-sample). There exist several ways to partition a data set into such a training and test set. The first way is chronologically. That is, the early data constitutes the training set, and the late data constitutes the test set.  \Cref{tab:literature-summary} indicates that most of the papers  follow this approach. However, some  studies violate this time structure in the data by choosing a different way to partition the data. Violations can be introduced by randomly partitioning the data into a training  and a test set or by  using a so-called `odd-even split.' 

Random partitioning breaks the time structure and introduces information leakage between the training set and the test set. When an ANN is trained on a training set constructed in such a way, the  error on the  test set underestimates the generalisation error of the ANN. \cite{yao2000option} and our companion paper \cite{ruf2019neural} provide more discussion on this point.  

 Some papers only work with independent draws from various distributions, and therefore do not involve any time series structure. Although these papers randomly partition the whole data set into a training and test set, no time structure is violated. Hence, in \Cref{tab:literature-summary}, we classify this approach as chronological partition. 

A related issue is the existence of time-inhomogeneity in financial data; in particular, volatility changes over time. When working with real data, some papers use a rolling window method to tackle this issue, especially when the time range is long and volatilities are not included as input features. Such papers include \cite{hutchinson1994nonparametric}, \cite{dugas2009incorporating}, and others. However, it remains an open question how big  window sizes need to be.

\subsection{Underlying assets and time span} \label{sec:undelryings}
	
Both simulation data and real data can be used to train an ANN for a specific problem. Simulation data is much easier to work with, since it is free of noise and sometimes a close-to-optimal solution is available as a benchmark, such as for the Black-Scholes and Heston models. For instance, \cite{le2001emulating}, \cite{morelli2004pricing}, and \cite{karatas2019supervised} investigate an ANN's performance on simulation data.  Most other papers  use either both  simulation  and real data or only real data. 
Options on S\&P500 have been studied by the largest number of papers, since they are the most liquidly traded options. Options on FTSE100 and S\&P100 have also been studied in several papers.  We refer to \Cref{tab:index-table} for a more complete list of all the underlyings being used.

Some papers focus on American option pricing and hedging. Underlyings for American options are usually individual stocks. Papers involving American options include \cite{kelly1994valuing}, \cite{chen1999pricing}, \cite{meissner2001capturing}, \cite{pires2004american}, \cite{pires2005american},  and \cite{amornwattana2007hybrid}.  As elaborated in Subsection~\ref{SS:4.3}, American options can also be priced differently by ANNs, via learning the value function or optimal stopping rule in a dynamic programming setting; see \cite{kohler2010pricing} and \cite{becker2019deep}.

\section{Recommended papers} \label{S:3}
Among the many papers of \Cref{tab:literature-summary}, we would like to highlight a few. Such a selection is clearly personal and subjective. Despite the subjective selection, we  believe that this list might serve as a good starting point to get an overview of this field.  We also provide  a Google Scholar citation count.\footnote{As of October 3, 2019.}  As  mentioned before, \Cref{tab:literature-summary} focuses only on  those papers that use ANNs to estimate option prices and related variables. Recently there have been many interesting and promising developments in the use of ANNs for calibration purposes or as computational tools. These papers are not included here, but Section~\ref{sec:other} provides some pointers to this literature.

Among the following highlighted papers, some are the first to propose innovative solutions. Others investigate the problem in a systematic way. 

\begin{itemize}
	\item \cite{hutchinson1994nonparametric} (\# citations: 749) is one of the first papers and the most highly cited one to use ANNs to estimate option prices. They introduce a methodology to evaluate the hedging performance over multiple periods, applied by many papers later on.
	
	\item \cite{lajbcygier1997improved} (\# citations:\footnote{This count includes the number of citations for \cite{lajbcygier1997improved2}.}  51) is one of the first papers that propose to learn the difference between model prices and  observed market option prices. 
	
	\item \cite{anders1998improving} (\# citations: 106) compare the  performance of ANNs and of the Black-Scholes benchmark when using different volatility estimates. 
	
	\item \cite{garcia2000pricing} (\# citations:\footnote{This count includes the number of citations for \cite{garcia1998option}.} 210) incorporate a homogeneity hint for the ANN. Hence, this is one of the first papers that embed financial domain knowledge into the construction of an ANN.
	
	\item \cite{carverhill2003alternative} (\# citations: 15) first propose an ANN that outputs hedging strategies directly, instead of option prices.

	\item \cite{bennell2004black} (\# citations: 83), \cite{chen2012pricing} (\#  citations: 12), and \cite{hahn2013option} (\# citations: 9) provide three extensive literature surveys.
	
	\item \cite{dugas2009incorporating} (\# citations:\footnote{This count includes the number of citations for \cite{dugas2001incorporating}.} 172) first design an ANN architecture that enforces no-arbitrage conditions such as convexity of option prices. 
		
	\item  \cite{andreou2010generalized} (\# citations: 19) combines an ANN with parametric models to learn 
	 functions that return implied model parameters. Such an  ANN essentially calibrates  parametric models.
	
	\item \cite{buehler2019deep} (\# citations: 23) develop a novel framework for hedging a portfolio of derivatives in the presence of market frictions, and allow  convex risk measures as loss functions. Their framework allows pricing and hedging without observing option prices.  

\end{itemize}

As this is a subjective selection, we also would like to highlight our companion paper \cite{ruf2019neural}, which provides a new benchmark based on delta-vega hedging and discusses data leakage issues.

\section{Related papers} \label{sec:other}

In the last few years, many novel techniques have been developed to apply ANNs to tasks arising in option pricing beyond the nonparametric estimation of  prices and hedging ratios. In this section we provide a few pointers to this rapidly developing literature.\footnote{At times it was not always clear cut to us whether a paper should be included in \Cref{tab:literature-summary} or in this section. For example,  the calibration papers of \Cref{SS:calibration} could have been put into  \Cref{tab:literature-summary} as mentioned in \Cref{sec:outputs}. Similarly, \cite{barucci1996no,barucci1997neural}, discussed in \Cref{SS:PDE}, learn the Black-Scholes model and hence could have been put into \Cref{tab:literature-summary}.}


\subsection{Calibration} \label{SS:calibration}
As already mentioned in \Cref{S:3}, \cite{andreou2010generalized} propose an ANN that returns implied model parameters. Hence, the ANN essentially calibrates parametric models. We observe a recent surge of the application of ANN to calibration. In this approach option prices are first mapped to a  parametric model, which is then used  to determine option prices. This approach can move the computationally heavy calibration off-line, thus significantly accelerating option pricing.  

\cite{abu2001financial} use neural networks to calibrate the Vasicek model with a consistency hint to produce valid parameters. More recently, \cite{hernandez2016calibration} uses an ANN to calibrate a single-factor Hull-White model. \cite{dimitroff2018volatility}, \cite{mcghee2018artificial} and \cite{liu2019neural} calibrate stochastic volatility models, and   \cite{stone2019calibrating} and \cite{bayer2019deep}\footnote{For more details, see also \cite{bayer2018deep} and \cite{horvath2019deep}.} calibrate rough volatility models.  \cite{itkin2019deep} highlights  some pitfalls in the existing approaches and proposes resolutions that improve both performance and accuracy of calibration.

Going the `indirect' way via first calibrating a model and then using it to determine the hedging ratio has at least two advantages. First, it provides additional interpretability as only the calibration step is replaced by an ANN. This can be important for a financial entity subject to regulatory requirements. Second, it provides an arguably strong tailor-made regularisation effect as it replaces a nonparametric estimation task  by the task of estimating a model with usually less than 5-10 parameters. 


\subsection{Solving partial differential equations} \label{SS:PDE}
The option pricing problem  sometimes involves solving a partial differential equation (PDE).  \cite{barucci1996no,barucci1997neural} use the Galerkin method and ANNs for solving the Black-Scholes PDE. \cite{weinan2017deep}, \cite{han2018solving}, and \cite{beck2019deep}  utilize ANNs to solve high-dimensional semilinear parabolic PDEs. They propose to reformulate the PDEs using backward stochastic differential equations, and the gradient of the unknown solutions is approximated by ANNs. Their numerical results suggest that the method is effective for a wide variety of (possibly high-dimensional) problems. One case study involves the pricing of European options on 100 defaultable underlying  assets.
There are several recent papers, such as \cite{henry2017deep}, \cite{sirignano2018dgm},  \cite{chan2019machine}, \cite{hure2019some} , \cite{jacquier2019deep}, and \cite{vidales2019unbiased}, who have developed this application of ANNs further.

\subsection{Approximating value functions in optimal control problems} \label{SS:4.3}
ANNs can be used to approximate value functions that appear in dynamic programming, for example arising in the American option pricing problem; see for example \cite{ye2019derivatives}. \cite{kohler2010pricing} use ANNs  to estimate continuation values for high-dimensional American option pricing. \cite{becker2019deep} use ANNs for optimal stopping problems by learning the optimal stopping rule from Monte Carlo samples. ANNs have also been proposed to approximate the value function of a dynamic program for real option pricing, see \cite{taudes1998real}.

In this context, we also mention \cite{fecamp2019risk}, who use an ANN  as a computational tool to solve  the pricing and hedging problem under market frictions such as transaction costs.


\subsection{Further work}
\cite{albanese2019xva} use an ANN to compute the conditional value-at-risk and expected shortfall necessary for certain XVA computations, by solving a quantile regression.

We would like to also mention \cite{halperin2017qlbs} and \cite{kolm2019dynamic} who suggest a reinforcement learning methodology to take market frictions into account for the option pricing task.  

Finally,   generative ANNs have been suggested recently as a non-parametric simulation tool for stock prices; see, for example,  \cite{henry2019generative}, \cite{kondratyev2019market}, and  \cite{wiese2019quant}. Such simulation engines could then be used for option pricing and hedging, a direction still to be explored systematically. Just after finishing this survey, \cite{wiese2019deep} proposed a generative ANN for option prices (instead of stock prices).
	
\section{Digression: regularisation techniques} \label{sec:regularization}
As the advance of hardware allows for  bigger ANNs to be built, regularization techniques have become more important as part of the ANN training. Such techniques  include $L^2$, dropout, early stopping, etc.; see \cite{ormoneit1999regularization}, \cite{genccay2001pricing}, \cite{genccay2003degree}, and \cite{liu2019pricing}. Complementing these universal regularisations, several papers embed financial domain knowledge into ANNs, either at the stage of  architecture design or training.  Let us here also mention the suggested feature design by \cite{lu2003data, lu2003digital}, who consider the pricing of exotic options and suggest to use digital option prices as features.

For the architecture design the following has been suggested:
		\begin{itemize}
		\item Homogeneity hint. \cite{garcia1998option,garcia2000pricing} incorporate a homogeneity hint by considering an ANN consisting of two parts, one controlled by  moneyness and the other controlled by time-to-maturity.
		\item Shape-restricted outputs.
		\cite{dugas2001incorporating,dugas2009incorporating}, \cite{lajbcygier2004improving},   \cite{yang2017gated},  \cite{huh2019pricing}, and \cite{zheng2019gated}  enforce certain no-arbitrage conditions such as monotonicity and convexity of the ANN pricing function by fixing an appropriate architecture. 		\end{itemize}
		
	At the training state the following techniques are being used:
	\begin{itemize}
		\item Data augmentation. \cite{yang2017gated}  and \cite{zheng2019gated} create additional synthetic options to help with the training of ANNs.
		\item Loss penalty. \cite{itkin2019deep}  and \cite{ackerer2019deep} add various penalty terms to the loss function. Those terms present 
		no-arbitrage conditions. For example,  parameter configurations that allow for calendar arbitrage are being penalised.
	\end{itemize}   

In the context of ANN training, we would like also to   mention   \cite{niranjan1996sequential}, \cite{de2000hierarchical, freitas2000sequential}, and \cite{palmer2019evolutionary}. These papers propose and examine novel training algorithms for ANNs and illustrate them in the context of option hedging; these algorithms include the extended Kalman filter, sequential Monte Carlo, and evolutionary algorithms.




\bibliography{reference}

\begin{thebibliography}{205}
\providecommand{\natexlab}[1]{#1}
\providecommand{\url}[1]{\texttt{#1}}
\expandafter\ifx\csname urlstyle\endcsname\relax
  \providecommand{\doi}[1]{doi: #1}\else
  \providecommand{\doi}{doi: \begingroup \urlstyle{rm}\Url}\fi

\bibitem[Abu-Mostafa(2001)]{abu2001financial}
Y.~S. Abu-Mostafa.
\newblock Financial model calibration using consistency hints.
\newblock \emph{IEEE transactions on neural networks}, 12\penalty0
  (4):\penalty0 791--808, 2001.

\bibitem[Ackerer et~al.(2019)Ackerer, Tagasovska, and Vatter]{ackerer2019deep}
D.~Ackerer, N.~Tagasovska, and T.~Vatter.
\newblock Deep smoothing of the implied volatility surface.
\newblock SSRN 3402942, 2019.

\bibitem[Ahmed and Swidler(1998)]{ahmed1998forecasting}
P.~Ahmed and S.~Swidler.
\newblock Forecasting properties of neural network generated volatility
  estimates.
\newblock In \emph{Decision Technologies for Computational Finance}, pages
  247--258, 1998.

\bibitem[Ahn et~al.(2012)Ahn, Kim, Oh, and Kim]{ahn2012applying}
J.~J. Ahn, D.~H. Kim, K.~J. Oh, and T.~Y. Kim.
\newblock Applying option {Greeks} to directional forecasting of implied
  volatility in the options market: {a}n intelligent approach.
\newblock \emph{Expert Systems with Applications}, 39\penalty0 (10):\penalty0
  9315--9322, 2012.

\bibitem[Albanese et~al.(2019)Albanese, Cr{\'e}pey, Hoskinson, and
  Saadeddine]{albanese2019xva}
C.~Albanese, S.~Cr{\'e}pey, R.~Hoskinson, and B.~Saadeddine.
\newblock {XVA} analysis from the balance sheet.
\newblock Retrieved on October 25, 2019 from
  \url{https://math.maths.univ-evry.fr/crepey/}, 2019.

\bibitem[Amilon(2003)]{amilon2003neural}
H.~Amilon.
\newblock A neural network versus {B}lack--{S}choles: a comparison of pricing
  and hedging performances.
\newblock \emph{Journal of Forecasting}, 22:\penalty0 317--335, 2003.

\bibitem[Amornwattana et~al.(2007)Amornwattana, Enke, and
  Dagli]{amornwattana2007hybrid}
S.~Amornwattana, D.~Enke, and C.~H. Dagli.
\newblock A hybrid option pricing model using a neural network for estimating
  volatility.
\newblock \emph{International Journal of General Systems}, 36\penalty0
  (5):\penalty0 558--573, 2007.

\bibitem[Anders et~al.(1998)Anders, Korn, and Schmitt]{anders1998improving}
U.~Anders, O.~Korn, and C.~Schmitt.
\newblock Improving the pricing of options: a neural network approach.
\newblock \emph{Journal of Forecasting}, 17\penalty0 (5-6):\penalty0 369--388,
  1998.

\bibitem[Andreou(2008)]{andreou2008parametric}
P.~C. Andreou.
\newblock \emph{{Parametric and Nonparametric Functional Estimation for Options
  Pricing with Applications in Hedging and Trading}}.
\newblock PhD thesis, University of Cyprus, 2008.

\bibitem[Andreou et~al.(2002)Andreou, Charalambous, and
  Martzoukos]{andreou2002critical}
P.~C. Andreou, C.~Charalambous, and S.~H. Martzoukos.
\newblock Critical assessment of option pricing methods using artificial neural
  networks.
\newblock In \emph{International Conference on Artificial Neural Networks},
  pages 1131--1136, 2002.

\bibitem[Andreou et~al.(2006)Andreou, Charalambous, and
  Martzoukos]{andreou2006robust}
P.~C. Andreou, C.~Charalambous, and S.~H. Martzoukos.
\newblock Robust artificial neural networks for pricing of {E}uropean options.
\newblock \emph{Computational Economics}, 27\penalty0 (2-3):\penalty0 329--351,
  2006.

\bibitem[Andreou et~al.(2008)Andreou, Charalambous, and
  Martzoukos]{andreou2008pricing}
P.~C. Andreou, C.~Charalambous, and S.~H. Martzoukos.
\newblock Pricing and trading {E}uropean options by combining artificial neural
  networks and parametric models with implied parameters.
\newblock \emph{European Journal of Operational Research}, 185\penalty0
  (3):\penalty0 1415--1433, 2008.

\bibitem[Andreou et~al.(2010)Andreou, Charalambous, and
  Martzoukos]{andreou2010generalized}
P.~C. Andreou, C.~Charalambous, and S.~H. Martzoukos.
\newblock Generalized parameter functions for option pricing.
\newblock \emph{Journal of Banking \& Finance}, 34\penalty0 (3):\penalty0
  633--646, 2010.

\bibitem[Avellaneda et~al.(1998)Avellaneda, Carelli, and
  Stella]{avellaneda1998following}
M.~Avellaneda, A.~Carelli, and F.~Stella.
\newblock Following the {B}ayes path to option pricing.
\newblock \emph{Journal of Computational Intelligence in Finance}, 1998.

\bibitem[Barone-Adesi and Whaley(1987)]{barone1987efficient}
G.~Barone-Adesi and R.~E. Whaley.
\newblock Efficient analytic approximation of {A}merican option values.
\newblock \emph{The Journal of Finance}, 42\penalty0 (2):\penalty0 301--320,
  1987.

\bibitem[Barucci et~al.(1996)Barucci, Cherubini, and Landi]{barucci1996no}
E.~Barucci, U.~Cherubini, and L.~Landi.
\newblock No-arbitrage asset pricing with neural networks under stochastic
  volatility.
\newblock In \emph{Neural Networks in Financial Engineering: Proceedings of the
  Third International Conference on Neural Networks in the Capital Markets},
  pages 3--16, 1996.

\bibitem[Barucci et~al.(1997)Barucci, Cherubini, and Landi]{barucci1997neural}
E.~Barucci, U.~Cherubini, and L.~Landi.
\newblock Neural networks for contingent claim pricing via the {Galerki}n
  method.
\newblock In \emph{Computational Approaches to Economic Problems}, pages
  127--141. 1997.

\bibitem[Barunikova and Barunik(2011)]{barunikova2011neural}
M.~Barunikova and J.~Barunik.
\newblock Neural networks as semiparametric option pricing tool.
\newblock \emph{Bulletin of the Czech Econometric Society}, 18, 2011.

\bibitem[Bates(1996)]{bates1996jumps}
D.~S. Bates.
\newblock Jumps and stochastic volatility: exchange rate processes implicit in
  {D}eutsche mark options.
\newblock \emph{The Review of Financial Studies}, 9\penalty0 (1):\penalty0
  69--107, 1996.

\bibitem[Bayer and Stemper(2018)]{bayer2018deep}
C.~Bayer and B.~Stemper.
\newblock Deep calibration of rough stochastic volatility models.
\newblock arXiv:1810.03399, 2018.

\bibitem[Bayer et~al.(2019)Bayer, Horvath, Muguruza, Stemper, and
  Tomas]{bayer2019deep}
C.~Bayer, B.~Horvath, A.~Muguruza, B.~Stemper, and M.~Tomas.
\newblock On deep calibration of (rough) stochastic volatility models.
\newblock arXiv:1908.08806, 2019.

\bibitem[Beck et~al.(2019)Beck, Becker, Cheridito, Jentzen, and
  Neufeld]{beck2019deep}
C.~Beck, S.~Becker, P.~Cheridito, A.~Jentzen, and A.~Neufeld.
\newblock Deep splitting method for parabolic {PDEs}.
\newblock arXiv:1907.03452, 2019.

\bibitem[Becker et~al.(2019)Becker, Cheridito, and Jentzen]{becker2019deep}
S.~Becker, P.~Cheridito, and A.~Jentzen.
\newblock Deep optimal stopping.
\newblock \emph{Journal of Machine Learning Research}, 20\penalty0
  (74):\penalty0 1--25, 2019.

\bibitem[Bennell and Sutcliffe(2004)]{bennell2004black}
J.~Bennell and C.~Sutcliffe.
\newblock Black--{S}choles versus artificial neural networks in pricing {FTSE}
  100 options.
\newblock \emph{Intelligent Systems in Accounting, Finance \& Management:
  International Journal}, 12\penalty0 (4):\penalty0 243--260, 2004.

\bibitem[Billio et~al.(2002)Billio, Corazza, and Gobbo]{billio2002option}
M.~Billio, M.~Corazza, and M.~Gobbo.
\newblock Option pricing via regime switching models and multilayer
  perceptrons: a comparative approach.
\newblock \emph{Rendiconti per gli Studi Economici Quantitativi},
  2002:\penalty0 39--59, 2002.

\bibitem[Blynski and Faseruk(2006)]{blynski2006comparison}
L.~Blynski and A.~Faseruk.
\newblock Comparison of the effectiveness of option price forecasting:
  {B}lack--{S}choles vs. simple and hybrid neural networks.
\newblock \emph{Journal of Financial Management \& Analysis}, 19\penalty0
  (2):\penalty0 46--58, 2006.

\bibitem[Boek et~al.(1995)Boek, Lajbcygier, Palaniswami, and
  Flitman]{boek1995hybrid}
C.~Boek, P.~Lajbcygier, M.~Palaniswami, and A.~Flitman.
\newblock A hybrid neural network approach to the pricing of options.
\newblock In \emph{Proceedings of ICNN'95--International Conference on Neural
  Networks}, volume~2, pages 813--817. IEEE, 1995.

\bibitem[Briegel and Tresp(2000)]{briegel2000dynamic}
T.~Briegel and V.~Tresp.
\newblock Dynamic neural regression models.
\newblock Retrieved on August 29, 2019 from
  \url{https://epub.ub.uni-muenchen.de/1571/}, 2000.

\bibitem[Buehler et~al.(2019{\natexlab{a}})Buehler, Gonon, Teichmann, and
  Wood]{buehler2019deep}
H.~Buehler, L.~Gonon, J.~Teichmann, and B.~Wood.
\newblock Deep hedging.
\newblock \emph{Quantitative Finance}, 19\penalty0 (8):\penalty0 1271--1291,
  2019{\natexlab{a}}.

\bibitem[Buehler et~al.(2019{\natexlab{b}})Buehler, Gonon, Teichmann, Wood,
  Mohan, and Kochems]{buehler2019deep-ri}
H.~Buehler, L.~Gonon, J.~Teichmann, B.~Wood, B.~Mohan, and J.~Kochems.
\newblock Deep hedging: hedging derivatives under generic market frictions
  using reinforcement learning.
\newblock SSRN 3355706, 2019{\natexlab{b}}.

\bibitem[Can and Fadda(2014)]{can2014nonparametric}
M.~Can and {\v S}.~Fadda.
\newblock A nonparametric approach to pricing options learning networks.
\newblock \emph{Southeast Europe Journal of Soft Computing}, 3\penalty0 (1),
  2014.

\bibitem[Cao et~al.(2019)Cao, Chen, and Hull]{cao2019neutral}
J.~Cao, J.~Chen, and J.~C. Hull.
\newblock A neural network approach to understanding implied volatility
  movements.
\newblock SSRN 3288067, 2019.

\bibitem[Carelli et~al.(2000)Carelli, Silani, and Stella]{carelli2000profiling}
A.~Carelli, S.~Silani, and F.~Stella.
\newblock Profiling neural networks for option pricing.
\newblock \emph{International Journal of Theoretical and Applied Finance},
  3\penalty0 (02):\penalty0 183--204, 2000.

\bibitem[Carr et~al.(2003)Carr, Geman, Madan, and Yor]{carr2003stochastic}
P.~Carr, H.~Geman, D.~B. Madan, and M.~Yor.
\newblock Stochastic volatility for {L}{\'e}vy processes.
\newblock \emph{Mathematical Finance}, 13\penalty0 (3):\penalty0 345--382,
  2003.

\bibitem[Carverhill and Cheuk(2003)]{carverhill2003alternative}
A.~P. Carverhill and T.~H. Cheuk.
\newblock Alternative neural network approach for option pricing and hedging.
\newblock SSRN 480562, 2003.

\bibitem[Chan-Wai-Nam et~al.(2019)Chan-Wai-Nam, Mikael, and
  Warin]{chan2019machine}
Q.~Chan-Wai-Nam, J.~Mikael, and X.~Warin.
\newblock Machine learning for semi linear {PDEs}.
\newblock \emph{Journal of Scientific Computing}, 79\penalty0 (3):\penalty0
  1667--1712, 2019.

\bibitem[Chang et~al.(2013)Chang, Wang, and Yeh]{chang2013forecasting}
T.-Y. Chang, Y.-H. Wang, and H.-Y. Yeh.
\newblock Forecasting of option prices using a neural network model.
\newblock \emph{Journal of Accounting, Finance \& Management Strategy},
  8\penalty0 (1):\penalty0 123--136, 2013.

\bibitem[Charalambous and Martzoukos(2005)]{charalambous2005hybrid}
C.~Charalambous and S.~H. Martzoukos.
\newblock Hybrid artificial neural networks for efficient valuation of real
  options and financial derivatives.
\newblock \emph{Computational Management Science}, 2\penalty0 (2):\penalty0
  155--161, 2005.

\bibitem[Chen and Sutcliffe(2012)]{chen2012pricing}
F.~Chen and C.~Sutcliffe.
\newblock Pricing and hedging short sterling options using neural networks.
\newblock \emph{Intelligent Systems in Accounting, Finance and Management},
  19\penalty0 (2):\penalty0 128--149, 2012.

\bibitem[Chen(2009)]{chen2009learning}
J.~Chen.
\newblock Learning the {Black}--{Scholes} formula via support vector machines.
\newblock In \emph{Recent Advances in Statistics Application and Related Areas,
  2nd Conference of the International Institute of Applied Statistics Studies},
  volume 1\&2, pages 756--760, 2009.

\bibitem[Chen and Lee(1999)]{chen1999pricing}
S.-H. Chen and W.-C. Lee.
\newblock Pricing call warrants with artificial neural networks: the case of
  the {T}aiwan derivative market.
\newblock In \emph{IJCNN'99. International Joint Conference on Neural Networks.
  Proceedings (Cat. No. 99CH36339)}, volume~6, pages 3877--3882. IEEE, 1999.

\bibitem[Chiu and Lin(2008)]{chiu2008exploring}
D.-Y. Chiu and C.-C. Lin.
\newblock Exploring internal mechanism of warrant in financial market with a
  hybrid approach.
\newblock \emph{Expert Systems with Applications}, 35\penalty0 (3):\penalty0
  1237--1245, 2008.

\bibitem[Choi et~al.(2004)Choi, Lee, Han, and Lee]{choi2004efficient}
H.-J. Choi, H.-S. Lee, G.-S. Han, and J.~Lee.
\newblock Efficient option pricing via a globally regularized neural network.
\newblock In \emph{International Symposium on Neural Networks}, pages 988--993,
  2004.

\bibitem[Corrado and Su(1996)]{corrado1996skewness}
C.~J. Corrado and T.~Su.
\newblock Skewness and kurtosis in {S\&P} 500 index returns implied by option
  prices.
\newblock \emph{Journal of Financial Research}, 19\penalty0 (2):\penalty0
  175--192, 1996.

\bibitem[Cox et~al.(1979)Cox, Ross, and Rubinstein]{cox1979option}
J.~C. Cox, S.~A. Ross, and M.~Rubinstein.
\newblock Option pricing: a simplified approach.
\newblock \emph{Journal of Financial Economics}, 7\penalty0 (3):\penalty0
  229--263, 1979.

\bibitem[Culkin and Das(2017)]{culkin2017machine}
R.~Culkin and S.~R. Das.
\newblock Machine learning in finance: the case of deep learning for option
  pricing.
\newblock \emph{Journal of Investment Management}, 15\penalty0 (4):\penalty0
  92--100, 2017.

\bibitem[Das and Padhy(2017)]{das2017new}
S.~P. Das and S.~Padhy.
\newblock A new hybrid parametric and machine learning model with homogeneity
  hint for {E}uropean-style index option pricing.
\newblock \emph{Neural Computing and Applications}, 28\penalty0 (12):\penalty0
  4061--4077, 2017.

\bibitem[de~Freitas et~al.(2000{\natexlab{a}})de~Freitas, Niranjan, and
  Gee]{de2000hierarchical}
J.~F.~G. de~Freitas, M.~Niranjan, and A.~H. Gee.
\newblock Hierarchical {B}ayesian models for regularization in sequential
  learning.
\newblock \emph{Neural Computation}, 12\penalty0 (4):\penalty0 933--953,
  2000{\natexlab{a}}.

\bibitem[de~Freitas et~al.(2000{\natexlab{b}})de~Freitas, Niranjan, Gee, and
  Doucet]{freitas2000sequential}
J.~F.~G. de~Freitas, M.~Niranjan, A.~H. Gee, and A.~Doucet.
\newblock Sequential {M}onte {C}arlo methods to train neural network models.
\newblock \emph{Neural Computation}, 12\penalty0 (4):\penalty0 955--993,
  2000{\natexlab{b}}.

\bibitem[Dimitroff et~al.(2018)Dimitroff, R{\"o}der, and
  Fries]{dimitroff2018volatility}
G.~Dimitroff, D.~R{\"o}der, and C.~Fries.
\newblock Volatility model calibration with convolutional neural networks.
\newblock SSRN 3252432, 2018.

\bibitem[Dindar and Marwala(2004)]{dindar2004option}
Z.~A. Dindar and T.~Marwala.
\newblock Option pricing using a committee of neural networks and optimized
  networks.
\newblock In \emph{2004 IEEE International Conference on Systems, Man and
  Cybernetics (IEEE Cat. No. 04CH37583)}, volume~1, pages 434--438. IEEE, 2004.

\bibitem[Dugas et~al.(2001)Dugas, Bengio, B{\'e}lisle, Nadeau, and
  Garcia]{dugas2001incorporating}
C.~Dugas, Y.~Bengio, F.~B{\'e}lisle, C.~Nadeau, and R.~Garcia.
\newblock Incorporating second-order functional knowledge for better option
  pricing.
\newblock In \emph{Advances in Neural Information Processing Systems}, pages
  472--478, 2001.

\bibitem[Dugas et~al.(2009)Dugas, Bengio, B{\'e}lisle, Nadeau, and
  Garcia]{dugas2009incorporating}
C.~Dugas, Y.~Bengio, F.~B{\'e}lisle, C.~Nadeau, and R.~Garcia.
\newblock Incorporating functional knowledge in neural networks.
\newblock \emph{Journal of Machine Learning Research}, 10\penalty0
  (Jun):\penalty0 1239--1262, 2009.

\bibitem[E et~al.(2017)E, Han, and Jentzen]{weinan2017deep}
W.~E, J.~Han, and A.~Jentzen.
\newblock Deep learning-based numerical methods for high-dimensional parabolic
  partial differential equations and backward stochastic differential
  equations.
\newblock \emph{Communications in Mathematics and Statistics}, 5\penalty0
  (4):\penalty0 349--380, 2017.

\bibitem[Fang and George(2017)]{fang2017application}
Z.~Fang and K.~George.
\newblock Application of machine learning: an analysis of {A}sian options
  pricing using neural network.
\newblock In \emph{2017 IEEE 14th International Conference on e-Business
  Engineering (ICEBE)}, pages 142--149. IEEE, 2017.

\bibitem[Fecamp et~al.(2019)Fecamp, Mikael, and Warin]{fecamp2019risk}
S.~Fecamp, J.~Mikael, and X.~Warin.
\newblock Risk management with machine-learning-based algorithms.
\newblock arXiv:1902.05287, 2019.

\bibitem[Ferguson and Green(2018)]{ferguson2018deeply}
R.~Ferguson and A.~Green.
\newblock Deeply learning derivatives.
\newblock SSRN 3244821, 2018.

\bibitem[Galindo-Flores(2000)]{galindo2000framework}
J.~Galindo-Flores.
\newblock A framework for comparative analysis of statistical and machine
  learning methods: an application to the {Black--Scholes} option pricing
  model.
\newblock \emph{Computational Finance 1999}, pages 635--660, 2000.

\bibitem[Garcia and Gen{\c{c}}ay(1998)]{garcia1998option}
R.~Garcia and R.~Gen{\c{c}}ay.
\newblock Option pricing with neural networks and a homogeneity hint.
\newblock In \emph{Decision Technologies for Computational Finance}, pages
  195--205, 1998.

\bibitem[Garcia and Gen{\c{c}}ay(2000)]{garcia2000pricing}
R.~Garcia and R.~Gen{\c{c}}ay.
\newblock Pricing and hedging derivative securities with neural networks and a
  homogeneity hint.
\newblock \emph{Journal of Econometrics}, 94\penalty0 (1-2):\penalty0 93--115,
  2000.

\bibitem[Gatheral and Jacquier(2014)]{gatheral2014arbitrage}
J.~Gatheral and A.~Jacquier.
\newblock Arbitrage-free {SVI} volatility surfaces.
\newblock \emph{Quantitative Finance}, 14\penalty0 (1):\penalty0 59--71, 2014.

\bibitem[Geigle(1999)]{geigle1999thesis}
D.~S. Geigle.
\newblock \emph{{An Artificial Neural Network Approach to the Valuation of
  Options and Forecasting of Volatility}}.
\newblock PhD thesis, Nova Southeastern University, 1999.

\bibitem[Geigle and Aronson(1999)]{geigle1999artificial}
D.~S. Geigle and J.~E. Aronson.
\newblock An artificial neural network approach to the valuation of options and
  forecasting of volatility.
\newblock \emph{Journal of Computational Intelligence in Finance}, 7\penalty0
  (6):\penalty0 19--25, 1999.

\bibitem[Gen{\c{c}}ay and Gibson(2007)]{genccay2007model}
R.~Gen{\c{c}}ay and R.~Gibson.
\newblock Model risk for {E}uropean-style stock index options.
\newblock \emph{IEEE Transactions on Neural Networks}, 18\penalty0
  (1):\penalty0 193--202, 2007.

\bibitem[Gen{\c{c}}ay and Qi(2001)]{genccay2001pricing}
R.~Gen{\c{c}}ay and M.~Qi.
\newblock Pricing and hedging derivative securities with neural networks:
  Bayesian regularization, early stopping, and bagging.
\newblock \emph{IEEE Transactions on Neural Networks}, 12\penalty0
  (4):\penalty0 726--734, 2001.

\bibitem[Gen{\c{c}}ay and Salih(2003)]{genccay2003degree}
R.~Gen{\c{c}}ay and A.~Salih.
\newblock Degree of mispricing with the {B}lack--{S}choles model and
  nonparametric cures.
\newblock \emph{Annals of Economics and Finance}, 4:\penalty0 73--101, 2003.

\bibitem[Ghaziri et~al.(2000)Ghaziri, Elfakhani, and Assi]{ghaziri2000neural}
H.~Ghaziri, S.~Elfakhani, and J.~Assi.
\newblock Neural networks approach to pricing options.
\newblock \emph{Neural Network World}, 10\penalty0 (1):\penalty0 271--277,
  2000.

\bibitem[Ghosn and Bengio(2002)]{ghosn2002multi}
J.~Ghosn and Y.~Bengio.
\newblock Multi-task learning for option pricing.
\newblock Retrieved on October 29, 2019 from
  \url{https://cirano.qc.ca/files/publications/2002s-53.pdf}, 2002.

\bibitem[Ghysels et~al.(1998)Ghysels, Patilea, Renault, and
  Torr{\`e}s]{ghysels1997nonparametric}
E.~Ghysels, V.~Patilea, {\'E}.~Renault, and O.~Torr{\`e}s.
\newblock Nonparametric methods and option pricing.
\newblock In D.~Hand and S.~Jacka, editors, \emph{Statistics in Finance},
  chapter~13, pages 261--282. John Wiley \& Sons, 1998.

\bibitem[Gradojevic and Kukolj(2011)]{gradojevic2011parametric}
N.~Gradojevic and D.~Kukolj.
\newblock Parametric option pricing: {a} divide-and-conquer approach.
\newblock \emph{Physica D: Nonlinear Phenomena}, 240\penalty0 (19):\penalty0
  1528--1535, 2011.

\bibitem[Gradojevic et~al.(2009)Gradojevic, Gen{\c{c}}ay, and
  Kukolj]{gradojevic2009option}
N.~Gradojevic, R.~Gen{\c{c}}ay, and D.~Kukolj.
\newblock Option pricing with modular neural networks.
\newblock \emph{IEEE Transactions on Neural Networks}, 20\penalty0
  (4):\penalty0 626--637, 2009.

\bibitem[Gregoriou et~al.(2007)Gregoriou, Healy, and
  Ioannidis]{gregoriou2007hedging}
A.~Gregoriou, J.~Healy, and C.~Ioannidis.
\newblock Hedging under the influence of transaction costs: an empirical
  investigation on {FTSE} 100 index options.
\newblock \emph{Journal of Futures Markets}, 27\penalty0 (5):\penalty0
  471--494, 2007.

\bibitem[Hahn(2013)]{hahn2013option}
J.~T. Hahn.
\newblock \emph{Option Pricing Using Artificial Neural Networks: An Australian
  Perspective}.
\newblock PhD thesis, Bond University, 2013.

\bibitem[Halperin(2017)]{halperin2017qlbs}
I.~Halperin.
\newblock {QLBS}: Q-learner in the {Black-Scholes (-Merton)} worlds.
\newblock arXiv:1712.04609, 2017.

\bibitem[Hamid and Habib(2005)]{hamid2005can}
S.~A. Hamid and A.~Habib.
\newblock Can neural networks learn the {B}lack-{S}choles model?: A simplified
  approach.
\newblock Retrieved on September 9, 2019 from
  \url{https://academicarchive.snhu.edu/bitstream/handle/10474/1662/cfs2005-01.pdf},
  2005.

\bibitem[Han et~al.(2018)Han, Jentzen, and E]{han2018solving}
J.~Han, A.~Jentzen, and W.~E.
\newblock Solving high-dimensional partial differential equations using deep
  learning.
\newblock \emph{Proceedings of the National Academy of Sciences}, 115\penalty0
  (34):\penalty0 8505--8510, 2018.

\bibitem[Hanke(1997)]{hanke1997neural}
M.~Hanke.
\newblock Neural network approximation of option pricing formulas for
  analytically intractable option pricing models.
\newblock \emph{Journal of Computational Intelligence in Finance}, 5\penalty0
  (5):\penalty0 20--27, 1997.

\bibitem[Hanke(1999{\natexlab{a}})]{hanke1999adaptive}
M.~Hanke.
\newblock Adaptive hybrid neural network option pricing.
\newblock \emph{Journal of Computational Intelligence in Finance}, 7\penalty0
  (5):\penalty0 33--39, 1999{\natexlab{a}}.

\bibitem[Hanke(1999{\natexlab{b}})]{hanke1999neural}
M.~Hanke.
\newblock Neural networks versus {Black-Scholes}: an empirical comparison of
  the pricing accuracy of two fundamentally different option pricing methods.
\newblock \emph{Journal of Computational Intelligence in Finance}, 5:\penalty0
  26--34, 1999{\natexlab{b}}.

\bibitem[Healy et~al.(2002)Healy, Dixon, Read, and Cai]{healy2002data}
J.~Healy, M.~Dixon, B.~Read, and F.~Cai.
\newblock A data-centric approach to understanding the pricing of financial
  options.
\newblock \emph{The European Physical Journal B}, 27\penalty0 (2):\penalty0
  219--227, 2002.

\bibitem[Healy et~al.(2003)Healy, Dixon, Read, and Cai]{healy2003confidence}
J.~V. Healy, M.~Dixon, B.~J. Read, and F.~F. Cai.
\newblock Confidence in data mining model predictions: a financial engineering
  application.
\newblock In \emph{IECON'03. 29th Annual Conference of the IEEE Industrial
  Electronics Society (IEEE Cat. No. 03CH37468)}, volume~2, pages 1926--1931.
  IEEE, 2003.

\bibitem[Healy et~al.(2004)Healy, Dixon, Read, and Cai]{healy2004confidence}
J.~V. Healy, M.~Dixon, B.~J. Read, and F.~F. Cai.
\newblock Confidence limits for data mining models of options prices.
\newblock \emph{Physica A: Statistical Mechanics and its Applications},
  344\penalty0 (1-2):\penalty0 162--167, 2004.

\bibitem[Healy et~al.(2007)Healy, Dixon, Read, and Cai]{healy2007non}
J.~V. Healy, M.~Dixon, B.~J. Read, and F.~F. Cai.
\newblock Non-parametric extraction of implied asset price distributions.
\newblock \emph{Physica A: Statistical Mechanics and its Applications},
  382\penalty0 (1):\penalty0 121--128, 2007.

\bibitem[Henry-Labord\`ere(2017)]{henry2017deep}
P.~Henry-Labord\`ere.
\newblock Deep primal-dual algorithm for {BSDEs}: {a}pplications of machine
  learning to {CVA} and {IM}.
\newblock SSRN 3071506, 2017.

\bibitem[Henry-Labord\`ere(2019)]{henry2019generative}
P.~Henry-Labord\`ere.
\newblock Generative models for financial data.
\newblock SSRN 3408007, 2019.

\bibitem[Hernandez(2017)]{hernandez2016calibration}
A.~Hernandez.
\newblock Model calibration with neural networks.
\newblock \emph{Risk Magazine}, pages 1--5, June 2017.

\bibitem[Herrmann and Narr(1997)]{herrmann1997neural}
R.~Herrmann and A.~Narr.
\newblock Neural networks and the evaluation of derivatives: some insights into
  the implied pricing mechanism of german stock index options.
\newblock Retrieved on August 29, 2019 from
  \url{http://finance.fbv.kit.edu/download/dp202.pdf}, 1997.

\bibitem[Heston(1993)]{heston1993closed}
S.~L. Heston.
\newblock A closed-form solution for options with stochastic volatility with
  applications to bond and currency options.
\newblock \emph{The Review of Financial Studies}, 6\penalty0 (2):\penalty0
  327--343, 1993.

\bibitem[Horvath et~al.(2019)Horvath, Muguruza, and Tomas]{horvath2019deep}
B.~Horvath, A.~Muguruza, and M.~Tomas.
\newblock Deep learning volatility.
\newblock arXiv:1901.09647, 2019.

\bibitem[Huang(2008)]{huang2008online}
S.-C. Huang.
\newblock Online option price forecasting by using unscented {Kalman} filters
  and support vector machines.
\newblock \emph{Expert Systems with Applications}, 34\penalty0 (4):\penalty0
  2819--2825, 2008.

\bibitem[Huang and Wu(2006)]{huang2006hybrid}
S.-C. Huang and T.-K. Wu.
\newblock A hybrid unscented {Kalman} filter and support vector machine model
  in option price forecasting.
\newblock In \emph{International Conference on Natural Computation}, pages
  303--312, 2006.

\bibitem[Huh(2019)]{huh2019pricing}
J.~Huh.
\newblock Pricing options with exponential {L}{\'e}vy neural network.
\newblock \emph{Expert Systems with Applications}, 127:\penalty0 128--140,
  2019.

\bibitem[Hull and White(2017)]{hull2017optimal}
J.~Hull and A.~White.
\newblock Optimal delta hedging for options.
\newblock \emph{Journal of Banking \& Finance}, 82:\penalty0 180--190, 2017.

\bibitem[Hur{\'e} et~al.(2019)Hur{\'e}, Pham, and Warin]{hure2019some}
C.~Hur{\'e}, H.~Pham, and X.~Warin.
\newblock Some machine learning schemes for high-dimensional nonlinear {PDEs}.
\newblock arXiv:1902.01599, 2019.

\bibitem[Hutchinson et~al.(1994)Hutchinson, Lo, and
  Poggio]{hutchinson1994nonparametric}
J.~M. Hutchinson, A.~W. Lo, and T.~Poggio.
\newblock A nonparametric approach to pricing and hedging derivative securities
  via learning networks.
\newblock \emph{The Journal of Finance}, 49\penalty0 (3):\penalty0 851--889,
  1994.

\bibitem[Itkin(2019)]{itkin2019deep}
A.~Itkin.
\newblock Deep learning calibration of option pricing models: some pitfalls and
  solutions.
\newblock arXiv:1906.03507, 2019.

\bibitem[Jacquier and Oumgari(2019)]{jacquier2019deep}
A.~Jacquier and M.~Oumgari.
\newblock Deep {PPDEs} for rough local stochastic volatility.
\newblock arXiv:1906.02551, 2019.

\bibitem[Jang and Lee(2019)]{jang2019generative}
H.~Jang and J.~Lee.
\newblock Generative {B}ayesian neural network model for risk-neutral pricing
  of {A}merican index options.
\newblock \emph{Quantitative Finance}, 19\penalty0 (4):\penalty0 587--603,
  2019.

\bibitem[Jung et~al.(2006)Jung, Kim, and Lee]{jung2006novel}
K.-H. Jung, H.-C. Kim, and J.~Lee.
\newblock A novel learning network for option pricing with confidence interval
  information.
\newblock In \emph{International Symposium on Neural Networks}, pages 491--497,
  2006.

\bibitem[Kakati(2005)]{kakati2005pricing}
M.~Kakati.
\newblock Pricing and hedging performances of artificial neural net in {Indian}
  stock option market.
\newblock \emph{The ICFAI Journal of Applied Finance}, 11\penalty0
  (1):\penalty0 62--73, 2005.

\bibitem[Kakati(2008)]{kakati2008option}
M.~Kakati.
\newblock Option pricing using {Adaptive Neuro-Fuzzy System (ANFIS)}.
\newblock \emph{ICFAI Journal of Derivatives Markets}, 5\penalty0 (2), 2008.

\bibitem[Karaali et~al.(1997)Karaali, Edelberg, and
  Higgins]{karaali1997modelling}
O.~Karaali, W.~Edelberg, and J.~Higgins.
\newblock Modelling volatility derivatives using neural networks.
\newblock In \emph{Proceedings of the IEEE/IAFE 1997 Computational Intelligence
  for Financial Engineering}, pages 280--286. IEEE, 1997.

\bibitem[Karatas et~al.(2019)Karatas, Oskoui, and Hirsa]{karatas2019supervised}
T.~Karatas, A.~Oskoui, and A.~Hirsa.
\newblock Supervised deep neural networks ({DNN}s) for pricing/calibration of
  vanilla/exotic options under various different processes.
\newblock arXiv:1902.05810, 2019.

\bibitem[Kelly(1994)]{kelly1994valuing}
D.~L. Kelly.
\newblock Valuing and hedging {A}merican put options using neural networks.
\newblock Retrieved on August 29, 2019 from
  \url{http://citeseerx.ist.psu.edu/viewdoc/download?doi=10.1.1.721.8497&rep=rep1&type=pdf},
  1994.

\bibitem[Kim et~al.(2006)Kim, Lee, and Lee]{kim2006local}
B.-H. Kim, D.~Lee, and J.~Lee.
\newblock Local volatility function approximation using reconstructed radial
  basis function networks.
\newblock In \emph{International Symposium on Neural Networks}, pages 524--530,
  2006.

\bibitem[Ko et~al.(2005)Ko, Lin, Chien, and Cheng]{ko2005hedging}
P.~Ko, P.~Lin, W.~Chien, and Y.~Cheng.
\newblock Hedging derivative securities based on the neural network coefficient
  model.
\newblock In \emph{Proceedings of the Eighth Joint Conference on Information
  Sciences}, pages 1163--1166, 2005.

\bibitem[Ko(2009)]{ko2009option}
P.-C. Ko.
\newblock Option valuation based on the neural regression model.
\newblock \emph{Expert Systems with Applications}, 36\penalty0 (1):\penalty0
  464--471, 2009.

\bibitem[Kohler et~al.(2010)Kohler, Krzy{\.z}ak, and
  Todorovic]{kohler2010pricing}
M.~Kohler, A.~Krzy{\.z}ak, and N.~Todorovic.
\newblock Pricing of high-dimensional {A}merican options by neural networks.
\newblock \emph{Mathematical Finance}, 20\penalty0 (3):\penalty0 383--410,
  2010.

\bibitem[Kolm and Ritter(2019)]{kolm2019dynamic}
P.~N. Kolm and G.~Ritter.
\newblock Dynamic replication and hedging: A reinforcement learning approach.
\newblock \emph{The Journal of Financial Data Science}, 1\penalty0
  (1):\penalty0 159--171, 2019.

\bibitem[Kondratyev and Schwarz(2019)]{kondratyev2019market}
A.~Kondratyev and C.~Schwarz.
\newblock The market generator.
\newblock SSRN 3384948, 2019.

\bibitem[Kou(2002)]{kou2002jump}
S.~G. Kou.
\newblock A jump-diffusion model for option pricing.
\newblock \emph{Management Science}, 48\penalty0 (8):\penalty0 1086--1101,
  2002.

\bibitem[Krause(1996)]{krause1996option}
J.~Krause.
\newblock Option pricing with neural networks.
\newblock In \emph{Proceedings of the Fourth European Congress on Intelligent
  Techniques and Soft Computing}, volume~3, pages 2206--2210, 1996.

\bibitem[Lachtermacher and Rodrigues~Gaspar(1996)]{lachtermacher1996neural}
G.~Lachtermacher and L.~Rodrigues~Gaspar.
\newblock Neural networks in derivative securities pricing forecasting in
  {Brazilian} capital markets.
\newblock In \emph{Neural Networks in Financial Engineering: Proceedings of the
  Third International Conference on Neural Networks in the Capital Markets},
  pages 92--97, 1996.

\bibitem[Lai(2014)]{lai2014comparison}
W.-N. Lai.
\newblock Comparison of methods to estimate option implied risk-neutral
  densities.
\newblock \emph{Quantitative Finance}, 14\penalty0 (10):\penalty0 1839--1855,
  2014.

\bibitem[Lajbcygier(2002)]{lajbcygier2002comparing}
P.~R. Lajbcygier.
\newblock Comparing conventional and artificial neural network models for the
  pricing of options.
\newblock In \emph{Neural Networks in Business: Techniques and Applications},
  pages 220--235. IGI Global, 2002.

\bibitem[Lajbcygier(2003)]{lajbcygier2003improving}
P.~R. Lajbcygier.
\newblock Improving option pricing with the product constrained hybrid neural
  network.
\newblock In \emph{Artificial Neural Networks and Neural Information
  Processing}, pages 615--621, 2003.

\bibitem[Lajbcygier(2004)]{lajbcygier2004improving}
P.~R. Lajbcygier.
\newblock Improving option pricing with the product constrained hybrid neural
  network.
\newblock \emph{IEEE Transactions on Neural Networks}, 15\penalty0
  (2):\penalty0 465--476, 2004.

\bibitem[Lajbcygier and Connor(1997{\natexlab{a}})]{lajbcygier1997improved}
P.~R. Lajbcygier and J.~T. Connor.
\newblock Improved option pricing using artificial neural networks and
  bootstrap methods.
\newblock \emph{International Journal of Neural Systems}, 8\penalty0
  (04):\penalty0 457--471, 1997{\natexlab{a}}.

\bibitem[Lajbcygier and Connor(1997{\natexlab{b}})]{lajbcygier1997improved2}
P.~R. Lajbcygier and J.~T. Connor.
\newblock Improved option pricing using bootstrap methods.
\newblock In \emph{Proceedings of International Conference on Neural Networks},
  volume~4, pages 2193--2197. IEEE, 1997{\natexlab{b}}.

\bibitem[Lajbcygier and Flitman(1996)]{lajbcygier1996comparison}
P.~R. Lajbcygier and A.~Flitman.
\newblock A comparison of non-parametric regression techniques for the pricing
  of options using an optimal implied volatility.
\newblock In \emph{Decision Technologies for Financial Engineering: Proceedings
  of the Fourth International Conference on Neural Networks in Capital
  Markets}, pages 201--213, 1996.

\bibitem[Lajbcygier et~al.(1996{\natexlab{a}})Lajbcygier, Boek, Flitman, and
  Palaniswami]{lajbcygier1996comparing}
P.~R. Lajbcygier, C.~Boek, A.~Flitman, and M.~Palaniswami.
\newblock Comparing conventional and artificial neural network models for the
  pricing of options on futures.
\newblock \emph{NeuroVe\$t Journal}, 4\penalty0 (5):\penalty0 16--24,
  1996{\natexlab{a}}.

\bibitem[Lajbcygier et~al.(1996{\natexlab{b}})Lajbcygier, Boek, Palaniswami,
  and Flitman]{lajbcygier1996neural}
P.~R. Lajbcygier, C.~Boek, M.~Palaniswami, and A.~Flitman.
\newblock Neural network pricing of all ordinaries {SPI} options on futures.
\newblock In \emph{Neural Networks in Financial Engineering: Proceedings of the
  Third International Conference on Neural Networks in the Capital Markets},
  1996{\natexlab{b}}.

\bibitem[Lajbcygier et~al.(1997)Lajbcygier, Flitman, Swan, and
  Hyndman]{lajbcygier1997pricing}
P.~R. Lajbcygier, A.~Flitman, A.~Swan, and R.~J. Hyndman.
\newblock The pricing and trading of options using a hybrid neural network
  model with historical volatility.
\newblock \emph{NeuroVe\$t Journal}, pages 27--41, 1997.

\bibitem[le~Roux and du~Toit(2001)]{le2001emulating}
L.~J. le~Roux and G.~S. du~Toit.
\newblock Emulating the {B}lack \& {S}choles model with a neural network.
\newblock \emph{Southern African Business Review}, 5\penalty0 (1):\penalty0
  54--57, 2001.

\bibitem[Leung et~al.(2009)Leung, Chen, and Mancha]{leung2009making}
M.~T. Leung, A.-S. Chen, and R.~Mancha.
\newblock Making trading decisions for financial-engineered derivatives: {a}
  novel ensemble of neural networks using information content.
\newblock \emph{Intelligent Systems in Accounting, Finance \& Management},
  16\penalty0 (4):\penalty0 257--277, 2009.

\bibitem[Liang et~al.(2006)Liang, Zhang, and Yang]{liang2006pricing}
X.~Liang, H.~Zhang, and J.~Yang.
\newblock Pricing options in {Hong Kong} market based on neural networks.
\newblock In \emph{International Conference on Neural Information Processing},
  pages 410--419, 2006.

\bibitem[Liang et~al.(2009)Liang, Zhang, Xiao, and Chen]{liang2009improving}
X.~Liang, H.~Zhang, J.~Xiao, and Y.~Chen.
\newblock Improving option price forecasts with neural networks and support
  vector regressions.
\newblock \emph{Neurocomputing}, 72\penalty0 (13-15):\penalty0 3055--3065,
  2009.

\bibitem[Lin and Yeh(2005)]{lin2005valuation}
C.-T. Lin and H.-Y. Yeh.
\newblock The valuation of {T}aiwan stock index option price---comparison of
  performances between {B}lack-{S}choles and neural network model.
\newblock \emph{Journal of Statistics and Management Systems}, 8\penalty0
  (2):\penalty0 355--367, 2005.

\bibitem[Liu and Huang(2016)]{liu2016performance}
D.~Liu and S.~Huang.
\newblock The performance of hybrid artificial neural network models for option
  pricing during financial crises.
\newblock \emph{Journal of Data Science}, 14\penalty0 (1):\penalty0 1--18,
  2016.

\bibitem[Liu and Zhang(2011)]{liu2011pricing}
D.~Liu and L.~Zhang.
\newblock Pricing {C}hinese warrants using artificial neural networks coupled
  with {M}arkov regime switching model.
\newblock \emph{International Journal of Financial Markets and Derivatives},
  2\penalty0 (4):\penalty0 314--330, 2011.

\bibitem[Liu(1996)]{liu1996option}
M.~Liu.
\newblock Option pricing with neural networks.
\newblock In \emph{Progress in Neural Information Processing}, volume~2, pages
  760--765, 1996.

\bibitem[Liu et~al.(2019{\natexlab{a}})Liu, Borovykh, Grzelak, and
  Oosterlee]{liu2019neural}
S.~Liu, A.~Borovykh, L.~A. Grzelak, and C.~W. Oosterlee.
\newblock A neural network-based framework for financial model calibration.
\newblock \emph{Journal of Mathematics in Industry}, Forthcoming,
  2019{\natexlab{a}}.

\bibitem[Liu et~al.(2019{\natexlab{b}})Liu, Oosterlee, and
  Bohte]{liu2019pricing}
S.~Liu, C.~W. Oosterlee, and S.~M. Bohte.
\newblock Pricing options and computing implied volatilities using neural
  networks.
\newblock \emph{Risks}, 7\penalty0 (1):\penalty0 1--22, 2019{\natexlab{b}}.

\bibitem[Liu et~al.(2019{\natexlab{c}})Liu, Cao, Ma, and Shen]{liu2019wavelet}
X.~Liu, Y.~Cao, C.~Ma, and L.~Shen.
\newblock Wavelet-based option pricing: an empirical study.
\newblock \emph{European Journal of Operational Research}, 272\penalty0
  (3):\penalty0 1132--1142, 2019{\natexlab{c}}.

\bibitem[Longstaff and Schwartz(2001)]{longstaff2001valuing}
F.~A. Longstaff and E.~S. Schwartz.
\newblock Valuing {A}merican options by simulation: a simple least-squares
  approach.
\newblock \emph{The Review of Financial Studies}, 14\penalty0 (1):\penalty0
  113--147, 2001.

\bibitem[Lu and Ohta(2003{\natexlab{a}})]{lu2003data}
J.~Lu and H.~Ohta.
\newblock A data and digital-contracts driven method for pricing complex
  derivatives.
\newblock \emph{Quantitative Finance}, 3\penalty0 (3):\penalty0 212--219,
  2003{\natexlab{a}}.

\bibitem[Lu and Ohta(2003{\natexlab{b}})]{lu2003digital}
J.~Lu and H.~Ohta.
\newblock Digital contracts-driven method for pricing complex derivatives.
\newblock \emph{Journal of the Operational Research Society}, 54\penalty0
  (9):\penalty0 1002--1010, 2003{\natexlab{b}}.

\bibitem[Ludwig(2015)]{ludwig2015robust}
M.~Ludwig.
\newblock Robust estimation of shape-constrained state price density surfaces.
\newblock \emph{The Journal of Derivatives}, 22\penalty0 (3):\penalty0 56--72,
  2015.

\bibitem[Madan et~al.(1998)Madan, Carr, and Chang]{madan1998variance}
D.~B. Madan, P.~P. Carr, and E.~C. Chang.
\newblock The variance {G}amma process and option pricing.
\newblock \emph{Review of Finance}, 2\penalty0 (1):\penalty0 79--105, 1998.

\bibitem[Malliaris and Salchenberger(1993{\natexlab{a}})]{malliaris1993beating}
M.~Malliaris and L.~Salchenberger.
\newblock Beating the best: a neural network challenges the {Black-Scholes}
  formula.
\newblock In \emph{Proceedings of 9th IEEE Conference on Artificial
  Intelligence for Applications}, pages 445--449. IEEE, 1993{\natexlab{a}}.

\bibitem[Malliaris and Salchenberger(1993{\natexlab{b}})]{malliaris1993neural}
M.~Malliaris and L.~Salchenberger.
\newblock A neural network model for estimating option prices.
\newblock \emph{Journal of Applied Intelligence}, 3\penalty0 (3):\penalty0
  193--206, 1993{\natexlab{b}}.

\bibitem[Malliaris and Salchenberger(1996)]{malliaris1996using}
M.~Malliaris and L.~Salchenberger.
\newblock Using neural networks to forecast the {S\&P}100 implied volatility.
\newblock \emph{Neurocomputing}, 10\penalty0 (2):\penalty0 183--195, 1996.

\bibitem[Martel et~al.(2009)Martel, Artiles, and
  Rodriguez]{martel2009financial}
C.~G. Martel, M.~D.~G. Artiles, and F.~F. Rodriguez.
\newblock A financial option pricing model based on learning algorithms.
\newblock In \emph{Proceedings of the World Multiconference on Applied
  Economics, Business and Development}, pages 153--157, 2009.

\bibitem[McGhee(2018)]{mcghee2018artificial}
W.~A. McGhee.
\newblock An artificial neural network representation of the {SABR} stochastic
  volatility model.
\newblock SSRN 3288882, 2018.

\bibitem[Meissner and Kawano(2001)]{meissner2001capturing}
G.~Meissner and N.~Kawano.
\newblock Capturing the volatility smile of options on high-tech stocks---a
  combined {GARCH}-neural network approach.
\newblock \emph{Journal of Economics and Finance}, 25\penalty0 (3):\penalty0
  276--292, 2001.

\bibitem[Miranda and Burgess(1995)]{miranda1995intraday}
F.~G. Miranda and N.~Burgess.
\newblock Intraday volatility forecasting for option pricing using a neural
  network approach.
\newblock In \emph{Proceedings of 1995 Conference on Computational Intelligence
  for Financial Engineering}, page~31. IEEE, 1995.

\bibitem[Mitra(2006)]{mitra2006improving}
S.~K. Mitra.
\newblock Improving accuracy of option price estimation using artificial neural
  networks.
\newblock SSRN 876881, 2006.

\bibitem[Mitra(2012)]{mitra2012option}
S.~K. Mitra.
\newblock An option pricing model that combines neural network approach and
  {B}lack {S}choles formula.
\newblock \emph{Global Journal of Computer Science and Technology}, 12\penalty0
  (4), 2012.

\bibitem[Montagna et~al.(2003)Montagna, Morelli, Nicrosini, Amato, and
  Farina]{montagna2003pricing}
G.~Montagna, M.~Morelli, O.~Nicrosini, P.~Amato, and M.~Farina.
\newblock Pricing derivatives by path integral and neural networks.
\newblock \emph{Physica A: Statistical Mechanics and its Applications},
  324\penalty0 (1-2):\penalty0 189--195, 2003.

\bibitem[Montesdeoca and Niranjan(2016)]{montesdeoca2016extending}
L.~Montesdeoca and M.~Niranjan.
\newblock Extending the feature set of a data-driven artificial neural network
  model of pricing financial options.
\newblock In \emph{2016 IEEE Symposium Series on Computational Intelligence
  (SSCI)}, pages 1--6. IEEE, 2016.

\bibitem[Morelli et~al.(2004)Morelli, Montagna, Nicrosini, Treccani, Farina,
  and Amato]{morelli2004pricing}
M.~J. Morelli, G.~Montagna, O.~Nicrosini, M.~Treccani, M.~Farina, and P.~Amato.
\newblock Pricing financial derivatives with neural networks.
\newblock \emph{Physica A: Statistical Mechanics and its Applications},
  338\penalty0 (1-2):\penalty0 160--165, 2004.

\bibitem[Mostafa(2011)]{mostafa2011application}
F.~Mostafa.
\newblock \emph{{Applications of Neural Networks in Market Risk}}.
\newblock PhD thesis, Curtin University, 2011.

\bibitem[Mostafa and Dillon(2008)]{mostafa2008neural}
F.~Mostafa and T.~Dillon.
\newblock A neural network approach to option pricing.
\newblock \emph{WIT Transactions on Information and Communication
  Technologies}, 41:\penalty0 71--85, 2008.

\bibitem[Niranjan(1996)]{niranjan1996sequential}
M.~Niranjan.
\newblock Sequential tracking in pricing financial options using model based
  and neural network approaches.
\newblock In \emph{Advances in Neural Information Processing Systems}, pages
  960--966, 1996.

\bibitem[Ormoneit(1999)]{ormoneit1999regularization}
D.~Ormoneit.
\newblock A regularization approach to continuous learning with an application
  to financial derivatives pricing.
\newblock \emph{Neural Networks}, 12\penalty0 (10):\penalty0 1405--1412, 1999.

\bibitem[Palmer(2019)]{palmer2019evolutionary}
S.~Palmer.
\newblock \emph{{Evolutionary Algorithms and Computational Methods for
  Derivatives Pricing}}.
\newblock PhD thesis, University College London, 2019.

\bibitem[Palmer and Gorse(2017)]{palmer2017pseudo}
S.~Palmer and D.~Gorse.
\newblock Pseudo-analytical solutions for stochastic options pricing using
  {M}onte {C}arlo simulation and breeding {PSO}-trained neural networks.
\newblock In \emph{European Symposium on Artificial Neural Networks,
  Computational Intelligence and Machine Learning}, pages 365--370, 2017.

\bibitem[Pande and Sahu(2006)]{pande2006new}
A.~Pande and R.~Sahu.
\newblock A new approach to volatility estimation and option price prediction
  for dividend paying stocks.
\newblock In \emph{WEHIA 2006--1st International Conference on Economic
  Sciences with Heterogeneous Interacting Agents; 15--17 June 2006, University
  of Bologna, Italy}, 2006.

\bibitem[Park et~al.(2014)Park, Kim, and Lee]{park2014parametric}
H.~Park, N.~Kim, and J.~Lee.
\newblock Parametric models and non-parametric machine learning models for
  predicting option prices: empirical comparison study over {KOSPI} 200 index
  options.
\newblock \emph{Expert Systems with Applications}, 41\penalty0 (11):\penalty0
  5227--5237, 2014.

\bibitem[Phani et~al.(2011)Phani, Chandra, and Raghav]{phani2011quest}
B.~Phani, B.~Chandra, and V.~Raghav.
\newblock Quest for efficient option pricing prediction model using machine
  learning techniques.
\newblock In \emph{The 2011 International Joint Conference on Neural Networks},
  pages 654--657. IEEE, 2011.

\bibitem[Pires and Marwala(2004{\natexlab{a}})]{pires2004american}
M.~M. Pires and T.~Marwala.
\newblock American option pricing using multi-layer perceptron and support
  vector machine.
\newblock In \emph{2004 IEEE International Conference on Systems, Man and
  Cybernetics (IEEE Cat. No. 04CH37583)}, volume~2, pages 1279--1285. IEEE,
  2004{\natexlab{a}}.

\bibitem[Pires and Marwala(2004{\natexlab{b}})]{pires2004option}
M.~M. Pires and T.~Marwala.
\newblock Option pricing using {B}ayesian neural networks.
\newblock In \emph{Fifteenth Annual Symposium of the Pattern Recognition
  Association of South Africa}, pages 161--166, 2004{\natexlab{b}}.

\bibitem[Pires and Marwala(2005)]{pires2005american}
M.~M. Pires and T.~Marwala.
\newblock American option pricing using {B}ayesian multi-layer perceptrons and
  {B}ayesian support vector machines.
\newblock In \emph{IEEE 3rd International Conference on Computational
  Cybernetics}, pages 219--224. IEEE, 2005.

\bibitem[Qi(1996)]{qi1996financial}
M.~Qi.
\newblock \emph{{Financial Applications of Generalized Nonlinear Nonparametric
  Econometric Methods (Artificial Neural Networks)}}.
\newblock PhD thesis, Ohio State University, 1996.

\bibitem[Qi and Maddala(1996)]{qi1996option}
M.~Qi and G.~Maddala.
\newblock Option pricing using artificial neural networks: the case of {S\&P
  }500 index call options.
\newblock In \emph{Neural Networks in Financial Engineering: Proceedings of the
  Third International Conference on Neural Networks in the Capital Markets},
  pages 78--91, 1996.

\bibitem[Quek et~al.(2008)Quek, Pasquier, and Kumar]{quek2008novel}
C.~Quek, M.~Pasquier, and N.~Kumar.
\newblock A novel recurrent neural network-based prediction system for option
  trading and hedging.
\newblock \emph{Applied Intelligence}, 29\penalty0 (2):\penalty0 138--151,
  2008.

\bibitem[Raberto et~al.(2000)Raberto, Cuniberti, Riani, Scales, Mainardi, and
  Servizi]{raberto2000learning}
M.~Raberto, G.~Cuniberti, M.~Riani, E.~Scales, F.~Mainardi, and G.~Servizi.
\newblock Learning short-option valuation in the presence of rare events.
\newblock \emph{International Journal of Theoretical and Applied Finance},
  3\penalty0 (03):\penalty0 563--564, 2000.

\bibitem[Ruf and Wang(2020)]{ruf2019neural}
J.~Ruf and W.~Wang.
\newblock Hedging with neural networks.
\newblock SSRN 3580132, 2020.

\bibitem[Saito and Jun(2000)]{saito2000neural}
S.~Saito and L.~Jun.
\newblock Neural network option pricing in connection with the {Black} and
  {Scholes} model.
\newblock In \emph{Proceedings of the Fifth Conference of the Asian Pacific
  Operations Research Society}, 2000.

\bibitem[Samur and Temur(2009)]{samur2009use}
Z.~I. Samur and G.~T. Temur.
\newblock The use of artificial neural network in option pricing: the case of
  {S\&P} 100 index options.
\newblock \emph{International Journal of Social, Behavioral, Educational,
  Economic, Business and Industrial Engineering}, 3\penalty0 (6):\penalty0
  644--649, 2009.

\bibitem[Saxena(2008)]{saxena2008valuation}
A.~Saxena.
\newblock Valuation of {S\&P} {CNX} {N}ifty options: comparison of
  {B}lack-{S}choles and hybrid {ANN} model.
\newblock In \emph{Proceedings SAS Global Forum}, 2008.

\bibitem[Schittenkopf and Dorffner(2001)]{schittenkopf2001risk}
C.~Schittenkopf and G.~Dorffner.
\newblock Risk-neutral density extraction from option prices: {i}mproved
  pricing with mixture density networks.
\newblock \emph{IEEE Transactions on Neural Networks}, 12\penalty0
  (4):\penalty0 716--725, 2001.

\bibitem[Shin and Ryu(2012)]{shin2012dynamic}
H.~J. Shin and J.~Ryu.
\newblock A dynamic hedging strategy for option transaction using artificial
  neural networks.
\newblock \emph{International Journal of Software Engineering and its
  Applications}, 6\penalty0 (4):\penalty0 111--116, 2012.

\bibitem[Sirignano and Spiliopoulos(2018)]{sirignano2018dgm}
J.~Sirignano and K.~Spiliopoulos.
\newblock {DGM}: a deep learning algorithm for solving partial differential
  equations.
\newblock \emph{Journal of Computational Physics}, 375:\penalty0 1339--1364,
  2018.

\bibitem[Stone(2019)]{stone2019calibrating}
H.~Stone.
\newblock Calibrating rough volatility models: a convolutional neural network
  approach.
\newblock \emph{Quantitative Finance}, pages 1--14, 2019.

\bibitem[Taudes et~al.(1998)Taudes, Natter, and Trcka]{taudes1998real}
A.~Taudes, M.~Natter, and M.~Trcka.
\newblock Real option valuation with neural networks.
\newblock \emph{Intelligent Systems in Accounting, Finance \& Management},
  7\penalty0 (1):\penalty0 43--52, 1998.

\bibitem[Teddy et~al.(2006)Teddy, Lai, and Quek]{teddy2006brain}
S.~D. Teddy, E.-K. Lai, and C.~Quek.
\newblock A brain-inspired cerebellar associative memory approach to option
  pricing and arbitrage trading.
\newblock In \emph{International Conference on Neural Information Processing},
  pages 370--379, 2006.

\bibitem[Teddy et~al.(2008)Teddy, Lai, and Quek]{teddy2008cerebellar}
S.~D. Teddy, E.-K. Lai, and C.~Quek.
\newblock A cerebellar associative memory approach to option pricing and
  arbitrage trading.
\newblock \emph{Neurocomputing}, 71\penalty0 (16-18):\penalty0 3303--3315,
  2008.

\bibitem[Thomaidis et~al.(2007)Thomaidis, Tzastoudis, and
  Dounias]{thomaidis2007comparison}
N.~S. Thomaidis, V.~S. Tzastoudis, and G.~Dounias.
\newblock A comparison of neural network model selection strategies for the
  pricing of {S\&P}500 stock index options.
\newblock \emph{International Journal on Artificial Intelligence Tools},
  16\penalty0 (06):\penalty0 1093--1113, 2007.

\bibitem[Tsaih(1999)]{tsaih1999sensitivity}
R.~Tsaih.
\newblock Sensitivity analysis, neural networks, and the finance.
\newblock In \emph{IJCNN'99. International Joint Conference on Neural Networks.
  Proceedings (Cat. No. 99CH36339)}, volume~6, pages 3830--3835. IEEE, 1999.

\bibitem[Tseng et~al.(2008)Tseng, Cheng, Wang, and Peng]{tseng2008artificial}
C.-H. Tseng, S.-T. Cheng, Y.-H. Wang, and J.-T. Peng.
\newblock Artificial neural network model of the hybrid {EGARCH} volatility of
  the {T}aiwan stock index option prices.
\newblock \emph{Physica A: Statistical Mechanics and its Applications},
  387\penalty0 (13):\penalty0 3192--3200, 2008.

\bibitem[Tung and Quek(2005)]{tung2005genso}
W.~L. Tung and C.~Quek.
\newblock {GenSo-OPATS}: a brain-inspired dynamically evolving option pricing
  model and arbitrage trading system.
\newblock In \emph{2005 IEEE Congress on Evolutionary Computation}, volume~3,
  pages 2429--2436. IEEE, 2005.

\bibitem[Tung and Quek(2011)]{tung2011financial}
W.~L. Tung and C.~Quek.
\newblock Financial volatility trading using a self-organising neural-fuzzy
  semantic network and option straddle-based approach.
\newblock \emph{Expert Systems with Applications}, 38\penalty0 (5):\penalty0
  4668--4688, 2011.

\bibitem[Tzastoudis et~al.(2006)Tzastoudis, Thomaidis, and
  Dounias]{tzastoudis2006improving}
V.~S. Tzastoudis, N.~S. Thomaidis, and G.~D. Dounias.
\newblock Improving neural network based option price forecasting.
\newblock In \emph{Hellenic Conference on Artificial Intelligence}, pages
  378--388, 2006.

\bibitem[Vidales et~al.(2019)Vidales, Siska, and Szpruch]{vidales2019unbiased}
M.~S. Vidales, D.~Siska, and L.~Szpruch.
\newblock Unbiased deep solvers for parametric {PDEs}.
\newblock arXiv:1810.05094, 2019.

\bibitem[von Spreckelsen et~al.(2014)von Spreckelsen, von Mettenheim, and
  Breitner]{von2014steps}
C.~von Spreckelsen, H.-J. von Mettenheim, and M.~H. Breitner.
\newblock Steps towards a high-frequency financial decision support system to
  pricing options on currency futures with neural networks.
\newblock \emph{International Journal of Applied Decision Sciences}, 7\penalty0
  (3):\penalty0 223--238, 2014.

\bibitem[Wang et~al.(2012)Wang, Lin, Huang, and Wu]{wang2012using}
C.-P. Wang, S.-H. Lin, H.-H. Huang, and P.-C. Wu.
\newblock Using neural network for forecasting {TXO} price under different
  volatility models.
\newblock \emph{Expert Systems with Applications}, 39\penalty0 (5):\penalty0
  5025--5032, 2012.

\bibitem[Wang(2006)]{wang2006dual}
H.-W. Wang.
\newblock Dual derivatives spreading and hedging with evolutionary data mining.
\newblock \emph{Journal of American Academy of Business}, 9:\penalty0 45--52,
  2006.

\bibitem[Wang(2011)]{wang2011pricing}
P.~Wang.
\newblock Pricing currency options with support vector regression and
  stochastic volatility model with jumps.
\newblock \emph{Expert Systems with Applications}, 38\penalty0 (1):\penalty0
  1--7, 2011.

\bibitem[Wang(2009{\natexlab{a}})]{wang2009nonlinear}
Y.-H. Wang.
\newblock Nonlinear neural network forecasting model for stock index option
  price: {h}ybrid {GJR--GARCH} approach.
\newblock \emph{Expert Systems with Applications}, 36\penalty0 (1):\penalty0
  564--570, 2009{\natexlab{a}}.

\bibitem[Wang(2009{\natexlab{b}})]{wang2009using}
Y.-H. Wang.
\newblock Using neural network to forecast stock index option price: {a} new
  hybrid {GARCH} approach.
\newblock \emph{Quality \& Quantity}, 43\penalty0 (5):\penalty0 833--843,
  2009{\natexlab{b}}.

\bibitem[White(2000)]{white2000pricing}
A.~White.
\newblock \emph{Pricing Options with Futures-Style Margining: a Genetic
  Adaptive Neural Network Approach}.
\newblock Garland Publishing, 2000.

\bibitem[White(1998)]{white1998genetic}
A.~J. White.
\newblock A genetic adaptive neural network approach to pricing options: {a}
  simulation analysis.
\newblock \emph{Journal of Computational Intelligence in Finance}, 6\penalty0
  (2):\penalty0 13--23, 1998.

\bibitem[Wiese et~al.(2019{\natexlab{a}})Wiese, Bai, Wood, and
  Buehler]{wiese2019deep}
M.~Wiese, L.~Bai, B.~Wood, and H.~Buehler.
\newblock Deep hedging: {l}earning to simulate equity option markets.
\newblock arXiv:1911.01700, 2019{\natexlab{a}}.

\bibitem[Wiese et~al.(2019{\natexlab{b}})Wiese, Knobloch, Korn, and
  Kretschmer]{wiese2019quant}
M.~Wiese, R.~Knobloch, R.~Korn, and P.~Kretschmer.
\newblock Quant {GANs}: deep generation of financial time series.
\newblock arXiv:1907.06673, 2019{\natexlab{b}}.

\bibitem[Xu et~al.(2004)Xu, Dixon, Eales, Cai, Read, and Healy]{xu2004barrier}
L.~Xu, M.~Dixon, B.~A. Eales, F.~F. Cai, B.~J. Read, and J.~V. Healy.
\newblock Barrier option pricing: {m}odelling with neural nets.
\newblock \emph{Physica A: Statistical Mechanics and its Applications},
  344\penalty0 (1-2):\penalty0 289--293, 2004.

\bibitem[Yang et~al.(2017)Yang, Zheng, and Hospedales]{yang2017gated}
Y.~Yang, Y.~Zheng, and T.~M. Hospedales.
\newblock Gated neural networks for option pricing: rationality by design.
\newblock In \emph{Association for the Advancement of Artificial Intelligence},
  pages 52--58, 2017.

\bibitem[Yao et~al.(2000)Yao, Li, and Tan]{yao2000option}
J.~Yao, Y.~Li, and C.~L. Tan.
\newblock Option price forecasting using neural networks.
\newblock \emph{Omega}, 28\penalty0 (4):\penalty0 455--466, 2000.

\bibitem[Ye and Zhang(2019)]{ye2019derivatives}
T.~Ye and L.~Zhang.
\newblock Derivatives pricing via machine learning.
\newblock SSRN 3352688, 2019.

\bibitem[Zapart(2002)]{zapart2002stochastic}
C.~Zapart.
\newblock Stochastic volatility options pricing with wavelets and artificial
  neural networks.
\newblock \emph{Quantitative Finance}, 2\penalty0 (6):\penalty0 487--495, 2002.

\bibitem[Zapart(2003{\natexlab{a}})]{zapart2003beyond}
C.~Zapart.
\newblock Beyond {B}lack--{S}choles: a neural networks-based approach to
  options pricing.
\newblock \emph{International Journal of Theoretical and Applied Finance},
  6\penalty0 (05):\penalty0 469--489, 2003{\natexlab{a}}.

\bibitem[Zapart(2003{\natexlab{b}})]{zapart2003statistical}
C.~Zapart.
\newblock Statistical arbitrage trading with wavelets and artificial neural
  networks.
\newblock In \emph{2003 IEEE International Conference on Computational
  Intelligence for Financial Engineering}, pages 429--435. IEEE,
  2003{\natexlab{b}}.

\bibitem[Zheng(2017)]{zheng2017machine}
Y.~Zheng.
\newblock \emph{{Machine Learning and Option Implied Information}}.
\newblock PhD thesis, Imperial College London, 2017.

\bibitem[Zheng et~al.(2019)Zheng, Yang, and Chen]{zheng2019gated}
Y.~Zheng, Y.~Yang, and B.~Chen.
\newblock Gated deep neural networks for implied volatility surfaces.
\newblock arXiv:1904.12834, 2019.

\bibitem[Zhou et~al.(2007)Zhou, Yang, and Han]{zhou2007nonparametric}
W.~Zhou, M.~Yang, and L.~Han.
\newblock A nonparametric approach to pricing convertible bond via neural
  network.
\newblock In \emph{Eighth ACIS International Conference on Software
  Engineering, Artificial Intelligence, Networking, and Parallel/Distributed
  Computing (SNPD 2007)}, volume~2, pages 564--569. IEEE, 2007.

\end{thebibliography}
\end{document}